# Calibrated Intervention and Containment of the COVID-19 Pandemic

Liang Tian[1,2,#], Xuefei Li[1,3,#], Fei Qi[1,3], Qian-Yuan Tang[1,4], Viola Tang[1,5], Jiang Liu[1,6], Zhiyuan Li[1,7], Xingye Cheng[1,2], Xuanxuan Li[1,8,9], Yingchen Shi[1,8,9], Haiguang Liu[1,9,10], Lei-Han Tang[1,2,9,*]

#Contributed equally

[1]COVID-19 Modelling Group, Hong Kong Baptist University, Kowloon, Hong Kong SAR, China, http://covid19group.hkbu.edu.hk
[2]Department of Physics and Institute of Computational and Theoretical Studies, Hong Kong Baptist University, Kowloon, Hong Kong SAR, China
[3]CAS Key Laboratory of Quantitative Engineering Biology, Shenzhen Institute of Synthetic Biology, Shenzhen Institutes of Advanced Technology, Shenzhen 518055, China
[4]Center for Complex Systems Biology, Universal Biology Institute, University of Tokyo, 113-0033 Tokyo, Japan
[5]Department of Information Systems, Business Statistics and Operations Management, Hong Kong University of Science and Technology, Hong Kong SAR, China
[6]Scarsdale, NY 10583, USA
[7]Center for Quantitative Biology, Peking University, Haidian, Beijing 100871, China
[8]Department of Engineering Physics, Tsinghua University, Haidian, Beijing 100084, China
[9]Complex Systems Division, Beijing Computational Science Research Center, Haidian, Beijing 100193, China
[10]Physics Department, Beijing Normal University, Haidian, Beijing 100875, China

*Correspondence: lhtang@hkbu.edu.hk, Lei-Han Tang, Department of Physics, Hong Kong Baptist University, Hong Kong SAR, China

# Abstract

Within a short period of time, COVID-19 grew into a world-wide pandemic. Transmission by pre-symptomatic and asymptomatic viral carriers rendered intervention and containment of the disease extremely challenging. Based on reported infection case studies, we construct an epidemiological model that focuses on transmission around the symptom onset. The model is calibrated against incubation period and pairwise transmission statistics during the initial outbreaks of the pandemic outside Wuhan with minimal non-pharmaceutical interventions. Mathematical treatment of the model yields explicit expressions for the size of latent and pre-symptomatic subpopulations during the exponential growth phase, with the local epidemic growth rate as input. We then explore reduction of the basic reproduction number $R_0$ through specific disease control measures such as contact tracing, testing, social distancing, wearing masks and sheltering in place. When these measures are implemented in combination, their effects on $R_0$ multiply. We also compare our model behaviour to the first wave of the COVID-19 spreading in various affected regions and highlight generic and less generic features of the pandemic development.

## Introduction

The Coronavirus Disease 2019 (COVID-19) is a new contagious disease caused by the novel coronavirus (SARS-COV-2) [1], which belongs to the genera of *betacoronavirus*, the same as the coronavirus that caused the SARS epidemic between 2002 and 2003 [2]. COVID-19 has spread to more than 200 countries/regions, with nearly 50 million confirmed cases and over 1.2 million lives claimed as of November 8, 2020 [3]. The outbreak has been declared a pandemic and a public health emergency of international concern [4].

As the specific symptoms of COVID-19 are now well-publicised, symptomatic transmissions are being contained in most countries. However, disease transmission by pre-symptomatic and asymptomatic viral carriers is seen to be extremely difficult to deal with due to its hidden nature [5]. Clinical data reveals that viral load becomes significant before the symptom onset [6–8]. Epidemiological investigations have identified clear cases of pre-symptomatic transmission soon after the initial outbreak [9–12]. Estimates vary greatly among experts on the percentage of total transmission due to this group of viral carriers, ranging from as low as 18% to over 50% [13–15]. An early model-based study by Ferretti *et al.* [16] suggested that pre-symptomatic transmission alone could yield a basic reproduction number $R_{0,\mathrm{p}} = 0.9$, close to the critical value of 1.0 that sustains epidemic growth. Under intense surveillance of the pandemic, pre-symptomatic and asymptomatic transmissions become the main focus in outbreak control [5].

While the actual viral shedding is influenced by many factors, patient viral load during the course of disease progression is more universal. This suggests a modelling approach that starts with clinical observations of symptom onset, and treats disease transmission as a dependent process that is further shaped by living and social conditions, including control measures to reduce physical contact. Following this strategy, we first introduce a model for an unprotected population and calibrate the model parameters against clinical case reports during the initial outbreak. Subsequently, we estimate the percentage reduction in the basic reproduction number (estimated to be around 3.87 at an exponential growth rate of 0.3/day) due to contact tracing, mask wearing and other measures, individually or in combination. Additionally, we present our findings against the epidemic development curves around the world to highlight the level of social mobilization required to contain COVID-19 spreading.

## Model

In epidemiological studies, the central quantity is the average number of infections per unit time $r(t)$ by a primary viral carrier who was infected at $t = 0$ [17,18]. In the case of COVID-19, disease transmission from a given individual peaks around his/her symptom onset time [7,8], as illustrated by the infectiousness curve shown in Fig. 1a (left panel). This property, when averaged over the population, gives an $r(t)$ (Fig. 1a, right panel) that closely resembles the symptom onset time distribution, which we denote by $p_{\mathrm{O}}(t)$ (Fig. 1a, middle panel). In fact, when the time window of transmission is narrowly centred around the symptom onset, we have approximately,

$$r(t) \approx R_{\mathrm{E}} p_{\mathrm{O}}(t + \theta_{\mathrm{S}}). \quad (1)$$

Equation (1) defines the primary model analysed in this study. The mean reproduction number $R_{\mathrm{E}}$ sets the overall level of disease transmission in the population, and equals the basic reproduction number $R_0$ when the infectious disease first breaks into a community. Its actual value could change over time due to factors such as the intervention and containment measures considered below. The shift parameter $\theta_S$ (Fig. 1a, right panel) accommodates changes in, e.g., isolation delays from long ($\theta_S < 0$) to short ($\theta_S > 0$).

Based on actual transmission data from case studies, we developed a more detailed model as presented in Supplementary Information (SI). As illustrated in Fig. 1a and 1b, the model takes into account the gap $\theta_P$ between the peak of the infectiousness curve and the clinically defined symptom onset. To incorporate this feature in a stochastic model for disease transmission among individuals, we divide the pre-symptomatic period into three phases, a non-infectious latent phase L, followed by an infectious pre-symptomatic phase A with two subphases $A_1$ and $A_2$ before and after the infectiousness peak. Starting from infection at $t = 0$, an individual first stays in the latent phase L. Transition to phase $A_1$ takes place at a duration-dependent rate $\alpha_L(t)$. Once in phase $A_1$, the individual is infectious with a daily transmission rate $\beta_A$. Progression to the next phase $A_2$ takes place at a rate $\alpha_A$. The duration of phase $A_2$ is fixed at $\theta_P$, after which symptoms develop and the person enters the symptomatic phase S. Upon entering $A_2$, the patient's disease transmission rate $\beta_B(\tau)$ weakens with the elapsed time $\tau = t - t_O + \theta_P$ to match the right-wing of the infectiousness curve.

Despite the non-Markovian nature of the above model, we are able to derive the following equation for epidemic development in a well-mixed community by focusing on the number of the infected individuals in phase $A_1$ (see SI, Sec. 1.1-1.3),

$$\frac{dA_1}{dt} = -\alpha_A A_1 + \int_{-\infty}^{t} K(t - t_1) A_1(t_1) dt_1, \tag{2}$$

where the kernel function is related to the mean reproduction rate $r(t)$ through the equation,

$$K(t) = \alpha_A r(t) + \frac{dr(t)}{dt}. \tag{3}$$

Under the assumption of constant transmission rate $\beta_A$ and exit rate $\alpha_A$, there is no need to keep track of the temporal profile of individuals inside the $A_1$ phase, which brings great simplification. Equations (2) and (3) can then be solved by performing the Laplace transform, and many analytical results follow. In this respect our model is equally tractable mathematically as the susceptible-exposed-infectious-recovered (SEIR) type models defined by a set of rate equations [19], while retaining the more general temporal structure for disease progression and transmission, which in turn allows for more precise evaluation of control measures that target specific subpopulations.

In the SI, we show that the mean reproduction number of the detailed model is given by $R_E = R_E^A + R_E^S$, with $R_E^A = \beta_A/\alpha_A + \int_0^{\theta_P} \beta_B(\tau) d\tau$ and $R_E^S = \int_{\theta_P}^{\infty} \beta_B(\tau) d\tau$ being reproduction numbers associated with pre-symptomatic and symptomatic transmissions, respectively. When the right wing of the infectiousness curve in Fig. 1a takes the form of an exponentially decaying function $\beta_B(\tau) = \beta_A e^{-\alpha_B \tau}$ with a sufficiently large decay rate $\alpha_B$, we recover Eq. (1) which was initially proposed on heuristic grounds. The shift parameter is given approximately by,

$$\theta_S \approx \theta_P - \frac{\alpha_A}{\alpha_B(\alpha_A + \alpha_B)}. \tag{4}$$

## Parameter Calibration and Basic Properties

### Incubation period

By combining three data sets [11,20,21] with a total of 347 infection cases outside the Hubei province in China, we estimated the incubation period statistics $p_{\mathrm{O}}(t)$ (see Fig. 2a). Due to the difficulty in identifying a precise date of infection, a window is assigned to the incubation period in each case. A rudimentary way to deal with the uncertainty is to treat all possible values inside the window as equally likely. This procedure yields a statistical distribution for each of the three data sets as well as the conglomerated one, as shown by symbols in Fig. 2a.

Alternatively, viewing the data as samples of a common underlying probability distribution, we estimated $p_{\mathrm{O}}(t)$ by likelihood maximization (see Methods and SI Sec. 2.1). Within the class of functions considered, the log-normal distribution combined with an exponential tail yields the largest likelihood value (Fig. 2a, red line). From day 6 onward, $p_{\mathrm{O}}(t)$ follows an exponential decay with a rate of $-0.31$/day, with a 95% confidence interval (CI) of ($-0.35$, $-0.27$) per day. We have also examined other values (from day 4 to day 8) for the switch. In all cases, exponential tail decay rates are found to be round $-0.31$/day (see SI Sec. 2.1).

### Infectiousness function

The infectiousness function is usually defined as how infectious an individual is in terms of time since infection. In view of the transmission characteristics of COVID-19, we quantify the normalized infectiousness around the symptom onset instead (see Methods). A data set of 77 pairwise transmissions in several eastern and southeastern Asian countries and regions during their initial COVID-19 outbreak was compiled by He *et al.* [7] We took 66 pairs to estimate the underlying probability distribution $p_{\mathrm{I}}(t)$ of the day of transmission $t$ with respect to the symptom onset of the primary case, with results shown in Fig. 2b. (See Methods and SI Sec. 2.2 for details.) Under a maximum likelihood estimation scheme, we considered three alternative forms for $p_{\mathrm{I}}(t)$. All have exponential tails far away from the transmission peak, but differ in the way the two wings are joined together in the peak region. In the first case, the two exponential tails join directly to produce a cusp in the middle. In the second case, a flat top of variable width is introduced. In the third case, the flat top is replaced by a parabolic cap to give a more rounded peak. It turns out that the cusp function, with its peak located at 0.68 days before the symptom onset, is the most probable for this data set (Fig. 2b). Decay rates for the left and right wings are given by 0.46/day and 0.54/day, respectively (see SI Sec. 2.2).

Due to its Markovian nature, the $\mathrm{A}_1$ phase has a duration that is also exponentially distributed. This gives rise to an exponential tail of the population-averaged infectiousness curve prior to entering the $\mathrm{A}_2$ phase. We therefore set the model parameters to $\alpha_{\mathrm{A}} = 0.46$/day, $\theta_{\mathrm{P}} = 0.68$ days, and $\beta_{\mathrm{B}}(\tau) = \beta_{\mathrm{A}} e^{-\alpha_{\mathrm{B}}\tau}$ with $\alpha_{\mathrm{B}} = 0.54$/day. These values were used in the numerical calculations presented below. The corresponding CIs are given in Table I.

### Serial interval

Xu *et al.* [22] compiled a database of 1407 COVID-19 transmission pairs outside the Hubei province in China between early January till mid-February 2020. Among them, 677 pairs have the symptom onset dates and social relationships of infector-infectees. A detailed analysis of the data set, stratified before, during and one week after the Wuhan lockdown on 23 January 2020, was carried out by Ali *et al.* [23] which showed reduction of the serial interval of symptom onsets by a factor of 3 over the five weeks. In Fig. 2c, we show the distribution of the serial interval data for the whole period (solid circles) and separately for the first (open squares) and last two weeks (open triangles) of the period. The red line gives the predicted serial interval distribution

$$p_{\mathrm{SI}}(t) = \int_{-\infty}^{t} p_{\mathrm{I}}(t')p_{\mathrm{O}}(t-t')\,dt' \tag{5}$$

using our estimated values for $p_{\mathrm{O}}(t)$ and $p_{\mathrm{I}}(t)$. While the overall agreement with the unstratified data is good especially on the positive side, it is also evident that serial intervals can be affected by factors such as the percentage of imported cases, the length of isolation delays, etc. which changed substantially before and after the Wuhan lockdown. As suggested in Ref. 23, their effect can be simulated with a shape function that masks $p_{\mathrm{I}}(t)$. For example, an imported case spent part of his/her infectious period outside the region where the data was collected, shifting $p_{\mathrm{SI}}(t)$ to the right. On the other hand, vigorous contact tracing shortens isolation delays significantly, which in turn shifts $p_{\mathrm{SI}}(t)$ to the left.

## Mean reproduction number

Under Eq. (1), the well-known Lotka–Euler estimating equation [24] yields,

$$R_{\mathrm{E}} = \frac{e^{-\lambda\theta_{\mathrm{S}}}}{\tilde{p}_{\mathrm{O}}(\lambda)}, \tag{6}$$

where $\tilde{p}_{\mathrm{O}}(\lambda) = \int_0^\infty p_{\mathrm{O}}(t)e^{-\lambda t}dt$ is the Laplace transform of $p_{\mathrm{O}}(t)$ (see SI Secs. 1.4 and 3.1). Using the estimated values above, we obtain from Eq. (6) the $R_{\mathrm{E}}$ versus $\lambda$ curve shown in Fig. 3a, which covers both the growth ($\lambda > 0$) and declining ($\lambda < 0$) phases of the epidemic. The slope of the curve at $R_{\mathrm{E}} = 1$ is given by $1/T_{\mathrm{g}}$, where $T_{\mathrm{g}}$ is the mean generation time and equals $\tau_{\mathrm{O}} - \theta_{\mathrm{S}} = 6.19$ days under Eq. (6). The intercept of the curve at $R_{\mathrm{E}} = 0$ gives an ultimate epidemic decay rate of $-0.31$/day when disease transmission comes to a complete halt.

To estimate the uncertainty in the computed $R_{\mathrm{E}}$-$\lambda$ curve, we performed bootstrap analysis of the data used to obtain $p_{\mathrm{O}}(t)$ and $p_{\mathrm{I}}(t)$. The detailed procedure is described in the SI (Sec. 2), with the result shown in Fig. 3a. At a growth rate of $\lambda = 0.3$/day, our estimated value of the basic reproduction number $R_0$ is 3.87 (95% CI [3.38, 4.48]).

## Composition of the infected population

As we demonstrate in the SI, the convolutional form of our main equation (2) enables many analytic results to be derived and evaluated with the calibrated parameters. Figure 3b shows the probabilities that a given individual is in one of the four phases on day *t* after infection, computed using the formula in Table S1 in the SI. The red line marks the boundary between the pre-symptomatic and symptomatic phases. The width of the orange-coloured region ($A_1$ phase), on the other hand, is proportional to $\alpha_{\mathrm{A}}^{-1} \approx 2$ days.

Figure 3c, obtained from the Laplace transforms of these curves, gives the percentage of the infected population in each of the four phases on a given day when the epidemic is growing at a rate $\lambda$. These curves allow for estimation of the hidden population in L, $A_1$ and $A_2$ phases from the knowledge of S in real-time. They form the basis for quantitative assessment of intervention measures. Note that at high growth rates, a larger percentage of the infected population is in the latent and pre-symptomatic phases, so that suppressing transmission by this group through, say mask wearing and social distancing, assumes a greater priority.

## Evaluation of Intervention Measures

### Testing and contact tracing

To break the transmission chain in the community, governments around the world have adopted two measures with varying levels of intensity: 1) testing and isolating infected individuals; and 2) tracing and quarantining contacts of infected individuals.

For testing control, persons in close proximity to a confirmed infection case are asked to undergo voluntary or mandatory testing for infection, and quarantined when the result is positive. From Fig. 3b we see that, if the test is conducted too close to the day of infection, the individual has a high probability to still be in the latent phase, hence the test result is likely to be negative. On the other hand, if the test is conducted too late, the person may have already infected others so that the reduction of $r(t)$ given by Eq. (1) is small. Therefore, there is an optimal window between the infection date and the test date, which we analyse in the SI. In Fig. 4a, we show the reduction of the basic reproduction number $R_0$ as a function of the reporting delay, assuming all suspected contacts are tested. At $R_0$ =3.87, if the results become available immediately after testing, the reduction of $R_0$ is shown as the blue curve, better than the testing outcomes with one day delay (red curve). The largest reduction is obtained when the test is performed 3 days after the contact. This corresponds to the day when the width of the orange plus dark blue region in Fig. 3b is the widest.

For contact tracing and quarantine, we show our results under the scenario that a fraction $q_{\mathrm{c}}$ of infectees are tracked down and quarantined within a time window $T_{\mathrm{trace}}$ since infection (Fig. 4b, blue line). This would bring the mean reproduction number $R_{\mathrm{E}}$ from $R_0 = 3.87$ to a value below 1 if full tracing and quarantine is executed within 6 days after contact. An 80% tracing efficiency shrinks the time window to 3-4 days for achieving the same effect (Fig. 4b, red line). Details can be found in Secs. 4.2 and 4.3 of the SI. The shaded areas on the plot, obtained from bootstrap analysis, show the range of the predicted reduction due to uncertainties in the incubation period estimation (see SI Sec. 2.1.3).

### Social distancing and mask wearing

Other than government-led intervention to break the transmission chain, individual-led efforts, including social-distancing, mask-wearing, frequent hand-washing, etc., can slow down or even stop the outbreak. Among them, radical shifts have taken place in people's attitudes towards population-wide mask wearing. It was enforced in most Asian countries since the initial phase of the outbreak, yet not adopted by EU and USA until June this year. As of August, community mask use was recommended or required by most major public health bodies [25,26]. However, despite multiple experiments performed on measuring the trapping efficacy of masks on viral particles at individual's level [27-30] the aggregate impact of mask wearing at the population level is not yet clearly quantified. Given the now established risk of pre-symptomatic transmission, and the dominant role of droplet-mediated COVID-19 infections [31], masks with relatively low efficacy for personal protection may nevertheless reduce the overall infections in a population [32]. Based on a previous study on influenza aerosols [33], we constructed a semi-quantitative model to show that mask-wearing reduces $r(t)$ and hence $R_{\mathrm{E}}$ by a factor $(1 - e \cdot p_{\mathrm{m}})^2$, where $e$ is the efficacy of trapping viral particles inside the mask, and $p_{\mathrm{m}}$ is the percentage of the mask-wearing population (see SI Sec. 4.4). According to this model, even for masks with intermediate efficacy ($e$ = 50%), population-wide mask-wearing at $p_{\mathrm{m}} = 98\%$ alone could bring down $R_{\mathrm{E}}$ from its basic value $R_0 = 3.87$ to 1.

When combined with contact tracing (Fig. 4c), the two effects multiply. Figure 4c shows a heatmap of the reduced $R_{\mathrm{E}}$ when contact tracing and isolation is completed within 5 days of infection. The solid black line indicates that the reduced $R_{\mathrm{E}}$ reaches 1. For example, the combination of tracing of close contacts at 60% efficiency within 5 days and 60% of the general public wearing masks achieves the same purpose. This target line can be reached with lower percentages when close contacts can be found within 2 days of possible infection (dash-dotted line), but the numbers need to be higher when the time frame is relaxed to 8 days (dashed line).

## Epidemic Development: Generic Features

We examined the temporal progression of COVID-19 outbreaks in different parts of the world using the data available from the Johns Hopkins CSSE Repository [34], with the aim to extract more universal aspects of the pandemic development in light of our model studies. Other than China, our exploration is limited to the initial stage of COVID-19 outbreaks till end of March 2020.

### Provincial outbreaks and containment in China

We focused on the daily confirmed cases from various provinces since the Wuhan lockdown on January 23, 2020. Broadly speaking, the ascending and descending curves follow very similar exponential laws, while the time it took to achieve the crossover was affected by the overall extent of the epidemic as well as occurrences of smaller outbreaks. From the data, we define three phases of the epidemic development.

Phase I is characterised by an exponential growth of the epidemic. In the first week after the Wuhan lockdown, nearly all provinces registered a growth rate of approximately 0.3/day (Fig. 5, region shaded in pink) in the newly confirmed cases. Reports indicate that most of the growth during this period was driven by imported cases from Hubei province, whose own growth continued at this rate for a longer period (Fig. 5a). The fraction of local infections during import-driven growth can be calculated and the result depends on the local value of $R_{\mathrm{E}}$ (see SI Sec. 4.5).

Phase II is a crossover phase where public policies on border control and local intervention measures become increasingly stringent. On a logarithmic scale, data from the most affected provinces (except Hubei) show consistent behaviour. Closer examination, however, reveals the presence of sporadic outbreaks. Well-documented examples include prison cases in Hubei, Shandong and Zhejiang provinces [35]. Overall, under the swift and forceful implementation of COVID-19 surveillance, turnaround of the epidemic in provinces other than Hubei was reached in about three weeks after the Wuhan lockdown. In Fig. 5b and the supplementary

Fig. S4, we present simulation results using our model, assuming a linear decrease of $R_E$ from a local value of 2.0 to zero over a period $T$, which indeed reproduces the data in Fig. 5. The more gradual change of $R_E$ assumed in our simulations can be interpreted as due to the progressive mobility control and isolation policies including additional lockdowns, which took place from February 4-10 [36,37], as well as allocation of massive resources by relevant authorities to conduct rigorous contact tracing and to rapidly expand isolation facilities for use by COVID-19 patients [38].

Phase III, or the final descent, occurred when the intervention measures essentially terminated transmission in the community. The few that re-emerged were quickly traced and contained. Within our model, the newly confirmed cases in this period are identified with the shrinking number of individuals moving from the latent to the symptomatic phase, as one moves along the time axis in Fig. 3b. Strikingly, the observed decay rate in this phase reached the maximum value of 0.31/day predicted by our model, including data from Hubei province shown in Fig. 5a. This observation indicates that the infected cases were isolated at extremely high efficiency. Interestingly, a similar decay in the daily new cases is seen on the cruise ship Diamond Princess (Fig. 5b).

**The first wave outside mainland China**

Figures 6a-c show the daily confirmed cases in selected countries and regions from late January till end of March 2020. Countries and regions in east Asia shown in Fig. 6a experienced the first wave sooner than the rest of the world, but the epidemic growth rate is much lower than other places due to the prevention measures in place such as border control and mask wearing by the general public. Despite these measures, South Korea documented a major outbreak in the second half of February that elevated the overall level of the epidemic in the country [39] (Fig. 6c). In Europe and the US, exponential growth of the pandemic, with a growth rate close to 0.3/day, were reported in a number of countries from the beginning of March onward (Fig. 6b), driven by local infections.

The surging pandemic triggered emergent response by public health authorities and governments at all levels. Towards the end of March, countries that adopted stringent intervention measures have seen a significant reduction of the pandemic growth rate (Fig. 6b). The government of Italy imposed a national quarantine on March 9 [40], after which growth in the number of newly confirmed cases slowed down [34]. On the other hand, South Korea implemented aggressive contact tracing and testing policies [41,42], enabling the country to bring the outbreak to a much-reduced level at $R_E \approx 1.0$.

In Fig. 6d we show the estimated epidemic growth rate $\lambda(t)$ against the cumulative number of confirmed cases $N(t)$ in five representative countries. We computed the growth rate from the local slope of the $\ln N(t)$ against $t$ curve, i.e., $\lambda(t) = \ln[N(t)/N(t-\Delta t)]/\Delta t$, using a time window $\Delta t = 3$ days. The interval between a few tens to a few thousands cumulative cases can be taken as the first phase of local outbreaks in these countries, where the estimated values of $\lambda(t)$ remain approximately stable. Three of the five countries exhibited growth rates of approximately 0.3/day during this period, while Iran and Japan assumed values above 0.4/day and around 0.1/day, respectively. It is evident that epidemic preparedness and cultural aspects significantly affected COVID-19 spreading in the local population, before government intervention and containment measures took effect. A more complete discussion of growth rates during the exponential phase in different countries and regions can be found in SI, Sec. 5.

## Summary and Discussions

We have succeeded in developing a directly calibratable model for COVID-19 transmission by both pre-symptomatic and symptomatic viral carriers. This was made possible by focusing on transmission around the symptom onset, which is a prominent feature of the disease. We then compartmentalised the pre-symptomatic period into three phases, and constructed a dynamical equation for the subpopulation in the $A_1$ phase where infectiousness reaches its maximum. The simplicity of the model allows for explicit mathematical expressions to be derived. Quantification of transmission risks at different stages of patient disease progression facilitates assessment of control measures, either to break the transmission chain or to reduce the overall level of social contacts in the community. For example, contact tracing in combination with mask wearing in public places, can have a strong and immediate effect in bringing down the epidemic growth. In reality, governments often take incremental steps in intervention measures to ease their impact on the economy and on people's livelihood. The quantitative treatment of epidemic control carried out in this study can serve as a reference in the decision-making process.

On a technical level, the modelling framework presented here is intuitive and flexible, and allows easy association of clinical features with population level pandemic development. This can be a significant advantage when the need arises to adapt the epidemic model to specific social environments and demographic composition. Our estimated incubation period distribution is in excellent agreement with other studies (see Table I for a comparison of key statistical features) and furthermore is not expected to change significantly over time. This places Eq. (1) as a convenient starting point for exploring temporal structures of the epidemic development. The shift parameter $\theta_S$ in the equation embodies, in an explicit form, changing patterns of disease transmission from symptomatic to the pre-symptomatic viral carriers, and hence can serve as an important index for epidemic control.

With regard to the quantitative predictions under specific intervention measures, the main uncertainty comes from estimation of their efficacy in reducing transmission from the infectious subpopulations identified in this study. As a baseline study, we estimated the infectiousness function $p_I(t)$ based on a relatively small data set of 66 transmission pairs which led to a sizable CI at 95% for its wings. This could improve as more carefully curated transmission cases during the initial outbreak become available. Response of the public to specific intervention measures is a complex topic that deserves extensive research in the future.

Finally, as with other epidemic models that assume a well-mixed population, our current modelling framework does not treat epidemic spreading in a heterogeneous population that exhibits complex spatio-temporal dynamics, nor does it consider significant differences in disease progression and transmission in different age groups. Some of the basic questions in COVID-19 epidemiological studies, such as whether pre-symptomatic spread constitutes a major contributor to disease transmission [43,44], cannot have definitive answers without considering these additional factors. In a large population, while individual outbreaks in specific communities may still follow the dynamics proposed here with suitable values of $R_E$, transmission across communities requires a separate treatment.

# Methods

## Key variables and parameters

We collect key variables and parameters of the compartmentalised model together with the estimated values in Table I for easy reference.

**Table I:** Key variables and parameters of the compartmentalised model.

| **Size variable** | **Subpopulation** | |
|---|---|---|
| L | Latent, infected but not infectious | |
| $A_1$ | Pre-symptomatic and infectious, constant transmission rate | |
| $A_2$ | Pre-symptomatic and infectious, decreasing transmission rate | |
| S | Symptomatic and infectious, diminishing transmission rate | |
| | | |
| **Epidemic characteristic** | **Definition (unit)** | |
| $R_{\mathrm{E}}$ | Mean reproduction number | |
| $\lambda$ | Exponential growth rate (per day) | |
| $r(t)$ | Mean reproduction rate since infection (per day) | |
| $p_{\mathrm{O}}(t)$ | Symptom onset time/Incubation period distribution | |
| $p_{\mathrm{I}}(t)$ | Distribution of infection time *t* in a transmission pair, measured from the symptom onset of the index patient | |
| $p_{\mathrm{SI}}(t)$ | Distribution of the delay time in the symptom onset of a transmission pair | |
| | | |
| **Parameter** | **Definition (unit)** | **Estimated value (95% CI)** |
| $t_{\mathrm{O}}$ | Symptom onset time/Incubation period (days) | Mean ($\tau_{\mathrm{O}}$): 6.04 (5.70, 6.37)[1]<br>Median: 4.60 (4.33, 4.88) [2]<br>Variance ($\sigma_{\mathrm{O}}$): 4.11 (3.77, 4.46) |
| $\gamma$ | Exponential decay rate of $p_{\mathrm{O}}(t)$ after 6 days (per day) | 0.31 (0.27, 0.35) |
| $\alpha_{\mathrm{L}}(t)$ | Transition rate from L to $A_1$ (per day) | Calibrated through $p_{\mathrm{O}}(t)$ (see SI Sec. 1 for details) |
| $\beta_{\mathrm{A}}$ | Transmission rate in $A_1$ phase (per day) | 0.97 (0.74, 1.27) at epidemic daily growth rate $\lambda = 0.3$/day |
| $\alpha_{\mathrm{A}}$ | Transition rate from $A_1$ to $A_2$ (per day) | 0.43 (0.32, 0.69) |
| $\beta_{\mathrm{B}}(t)$ | Transmission rate in $A_2$ + S (per day) | $\beta_{\mathrm{B}}(t) = \beta_{\mathrm{A}} e^{-\alpha_{\mathrm{B}} t}$ |
| $\alpha_{\mathrm{B}}$ | Decay rate of infectiousness in $A_2$ + S (per day) | 0.54 (0.48, 0.65) |
| $R_0$ | Basic reproduction number | 3.87 (3.38, 4.48) at epidemic growth rate $\lambda = 0.3$/day |
| $\theta_{\mathrm{P}}$ | Duration of $A_2$ phase (days) | 0.68 (0.12, 1.02) |
| $\theta_{\mathrm{S}}$ | Reproduction shift parameter (days) | -0.15 (-0.60, 0.23) |

[1] Mean incubation period was estimated at 5.95 days (95% CI 4.94, 7.11) and 6.4 days (95% CI 5.6, 7.7) in Refs. 21 and 45, respectively.
[2] Median incubation period was estimated at 5.1 days (95% CI 4.5, 5.8) in Ref. 46.

**Maximum likelihood estimation and uncertainty/sensitivity analysis**

Incubation period distribution – We analysed incubation periods of a total of $N = 347$ cases by combining three datasets [11,20,21]. For most cases, the infection date can only be assigned to a time window of more than one day. Therefore, the actual incubation period falls between $\text{IPl}_i$ and $\text{IPu}_i$, $i = 1, ..., N$, where $\text{IPl}_i$ and $\text{IPu}_i$ are the lower and upper bounds for case *i*. We perform maximum likelihood estimation of the underlying symptom onset time distribution $p_\text{O}(t)$, following a scheme proposed by Reich *et al.* [47] Considering the exponential tail observed in the real data, we write,

$$p_\text{O}(\theta, t) = \begin{cases} Ap_\text{left}(t) & \text{for } t \le t_\text{e} \\ Ap_\text{left}(t_\text{e})e^{-\gamma(t-t_\text{e})} & \text{for } t \ge t_\text{e} \end{cases}$$

where $A$ is the normalisation factor. Transition to the exponential decay (with rate $\gamma$) takes place at $t_\text{e}$. Following common practice in the epidemiological literature, we take $p_\text{left}(t)$ to be a truncated log-normal or Weibull distribution with 2 parameters in each case,

$$p_\text{left}(t) = \begin{cases} \dfrac{1}{t\sigma\sqrt{2\pi}} \exp\left[-\dfrac{(\ln t - \mu)^2}{2\sigma^2}\right] & \text{log-normal} \\ \dfrac{k}{\lambda}\left(\dfrac{t}{\lambda}\right)^{k-1} \exp\left[-\left(\dfrac{t}{\lambda}\right)^k\right] & \text{Weibull} \end{cases}$$

Continuity of derivatives at $t_\text{e}$ yields,

$$\gamma = -\frac{p'_\text{left}(t_\text{e})}{p_\text{left}(t_\text{e})}.$$

Thus we are left with a set of 3 independent parameters. To estimate these parameters from the data, we consider the likelihood function,

$$L(\theta; \mathbf{IP}) = \prod_{i=1}^{N} L(\theta; \text{IPl}_i, \text{IPu}_i)$$

with

$$L_i = L(\theta; \text{IPl}_i, \text{IPu}_i) = \int_{\text{IPl}_i - 0.5}^{\text{IPu}_i + 0.5} p_\text{O}(\theta, t)dt$$

We performed optimization and sensitivity analyses by scanning $t_\text{e}$ values from 4 to 8, and infinity for log-normal distribution and from 3 to 7, and infinity for Weibull distribution. The best estimate is obtained when $p_\text{left}(t)$ is a truncated log-normal distribution with $t_\text{e} = 6$ (see SI Sec. 2.1 for details).

We also performed bootstrap analysis to determine uncertainties in the estimated $p_\text{O}(t)$. This is done by generating 1000 re-sampled copies of the original dataset with 347 cases. The maximum likelihood estimation of $p_\text{O}(t)$ is then performed for each of the re-sampled copy. The 95% CIs were obtained from the 1,000 replications (see Table I).

Infectiousness profile – Disease transmission is quantified by the infectiousness function $p_\text{I}(t)$, which gives the probability density function for pairwise transmission at time *t* since the symptom onset of the infector. We infer $p_\text{I}(t)$ by maximum likelihood estimation, using the infector-infectee pairs published by He *et al.* [7]. In this dataset, the infectee exposure windows were documented in addition to the symptom onset dates of both infectors and infectees (77 pairs in total). Among them, 66 pairs have a unique symptom onset date (see Source data), which are used here.

Given the general form and the limited temporal resolution of the dataset, we adopted simple exponentials for the two wings of the infectiousness function joined in the middle by a cap function,

$$p_{\mathrm{I}}(\theta,t)=\begin{cases}Af(t_{\mathrm{A}})e^{\alpha_{\mathrm{A}}(t-t_{\mathrm{A}})} & t\le t_{\mathrm{A}}\\ Af(t) & t_{\mathrm{A}}\le t\le t_{\mathrm{B}}\\ Af(t_{\mathrm{B}})e^{-\alpha_{\mathrm{B}}(t-t_{\mathrm{B}})} & t\ge t_{\mathrm{B}}\end{cases}$$

where $A$ is the normalisation factor. The infectiousness function transits to the left exponential tail at $t_{\mathrm{A}}$ and to the right exponential tail at $t_{\mathrm{B}}$. Between $t_{\mathrm{A}}$ and $t_{\mathrm{B}}$, it takes the form of $f(t)$. We consider three different forms of $f(t)$:

- Model 1: $f(t)=1$ and $t_{\mathrm{A}}=t_{\mathrm{B}}=t_{\mathrm{P}}$ (two exponential tails directly join at $t_{\mathrm{P}}$); Independent parameters $\theta=(\alpha_{\mathrm{A}},\alpha_{\mathrm{B}},t_{\mathrm{P}})$.
- Model 2: $f(t)=1$ and $t_{\mathrm{A}}<t_{\mathrm{B}}$ (two exponential tails with a flat cap of length $\epsilon=t_{\mathrm{B}}-t_{\mathrm{A}}$, centred at $t_{\mathrm{P}}$); Independent parameters $\theta=(\alpha_{\mathrm{A}},\alpha_{\mathrm{B}},\epsilon,t_{\mathrm{P}})$.
- Model 3: $f(t)=[1-\chi(t-t_{\mathrm{P}})^2]$ and $t_{\mathrm{A}}<t_{\mathrm{B}}$ (two exponential tails with a rounded cap peaked at $t_{\mathrm{P}}$, whose shape is characterized by $\chi$); Independent parameters $\theta=(\alpha_{\mathrm{A}},\alpha_{\mathrm{B}},\chi,t_{\mathrm{P}})$ ($t_{\mathrm{A}}$ and $t_{\mathrm{B}}$ are determined by the smoothness condition).

We perform maximum likelihood estimations using the dataset mentioned above, where each transmission pair $i$ is associated with an exposure window $\mathrm{W}_i=[\mathrm{Wl}_i,\mathrm{Wu}_i]$ relative to the symptom onset of the infector. The likelihood function is constructed as follows:

$$L(\theta;\mathbf{W})=\prod_{i=1}^{N}L(\theta;\mathrm{Wl}_i,\mathrm{Wu}_i),$$

where

$$L_i=L(\theta;\mathrm{Wl}_i,\mathrm{Wu}_i)=\int_{\mathrm{Wl}_i-0.5}^{\mathrm{Wu}_i+0.5}p_{\mathrm{I}}(\theta,t)dt$$

Sensitivity analysis is performed at a set of values for $\epsilon$ (Model 2) and $\chi$ (Model 3), respectively. In both cases, the best estimate degenerates into Model 1 (see SI Sec. 2.2 for details).

The uncertainty in the estimated $p_{\mathrm{I}}(t)$ is determined through bootstrapping with 1,000 replications, with which the 95% CIs were obtained (see Table I).

# Data availability

All the data used in this work were previously published and publicly available (Source data).

# Code availability

Code to carry out analyses is publicly available on Github: https://github.com/hkbu-covid19group/Calib-covid19.

# Acknowledgments

The work is supported in part by the NSFC under Grant Nos. 11635002 and U1930402, and by the Research Grants Council of the Hong Kong Special Administrative Region (HKSAR) under Grant HKBU 12324716.

# Conflict of Interest Statement

Authors declare no conflict of interest.

# Disclaimer

Any views expressed by Jiang Liu and Viola Tang are not as a representative speaking for or on behalf of his/her employer, nor do they represent his/her employer's positions, strategies or opinions.

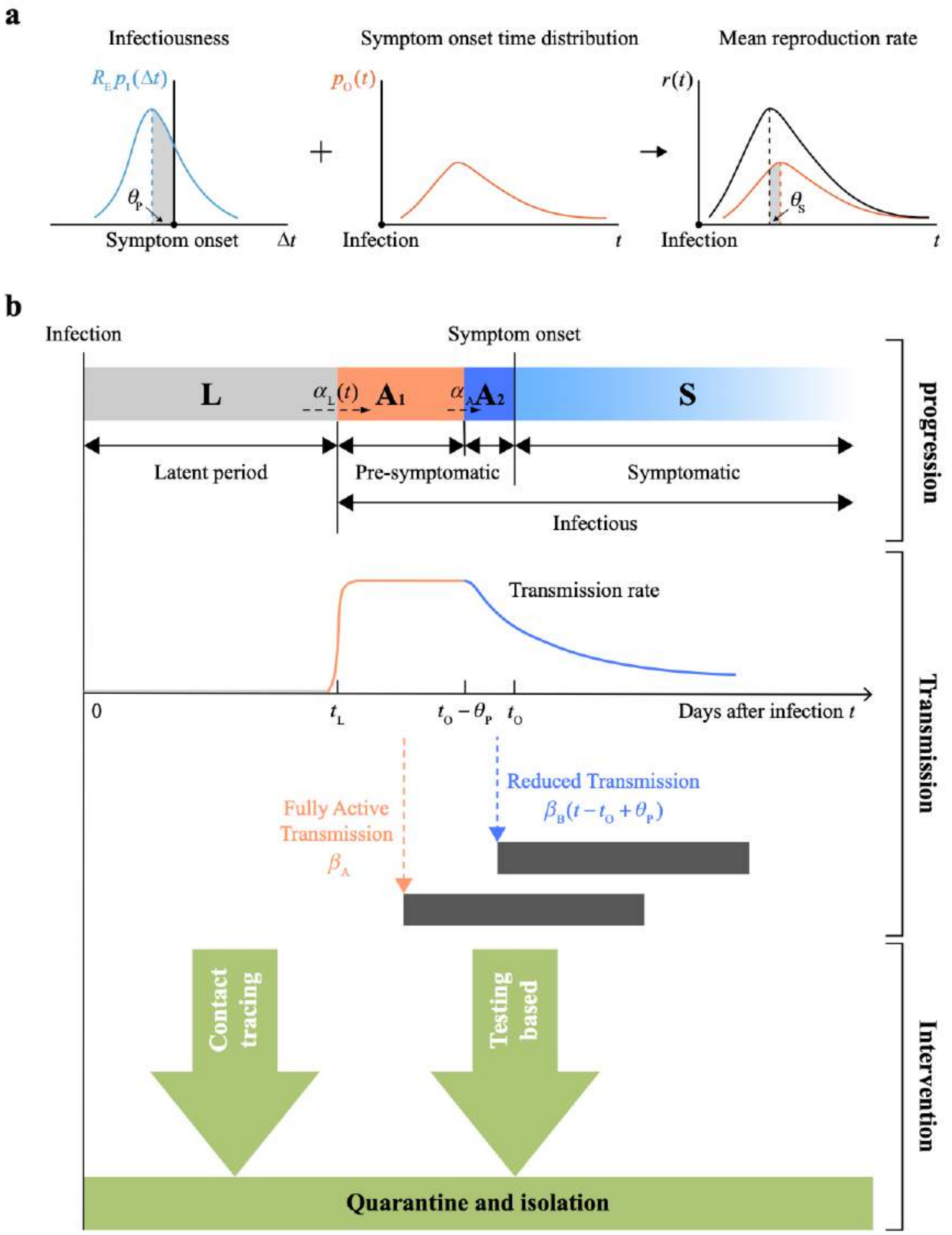

**Figure 1. A stochastic model for COVID-19 disease progression, transmission and intervention. a,** The mean reproduction rate $r(t)$ (black curve) of a patient since infection at $t = 0$ is expressed as a convolution of the symptom onset time distribution $p_O(t)$ (red curve) and the infectiousness curve $R_E p_I(\Delta t)$ (blue curve), where $\Delta t$ is measured from the symptom onset. The mean reproduction number $R_E$ sets the overall level of the epidemic. The peak of the normalised infectiousness function $p_I(\Delta t)$ is shifted from the symptom onset by an amount $\theta_P$, which takes a positive value on the pre-symptomatic side. The peak of the mean reproduction rate $r(t)$ is shifted from the peak of the symptom onset time distribution $p_O(t)$ by $\theta_S$. **b,** A compartmentalized model. A person infected at time $t = 0$ first goes through a non-infectious latent phase (L) until $t_L$, followed by an infectious period that spans across symptom onset at $t_O$. In the pre-symptomatic phase A, the person is infectious without symptoms. The A phase is further split into two subphases, $A_1$ with a constant transmission rate (orange region) and $A_2$ with a declining transmission rate (blue region). At the symptom onset time $t_O$, the person enters the S phase, and continues to be infectious (light blue region). Contact tracing brings an infected person out of the transmission cycle at the point of isolation, while testing does so only when the result is positive.

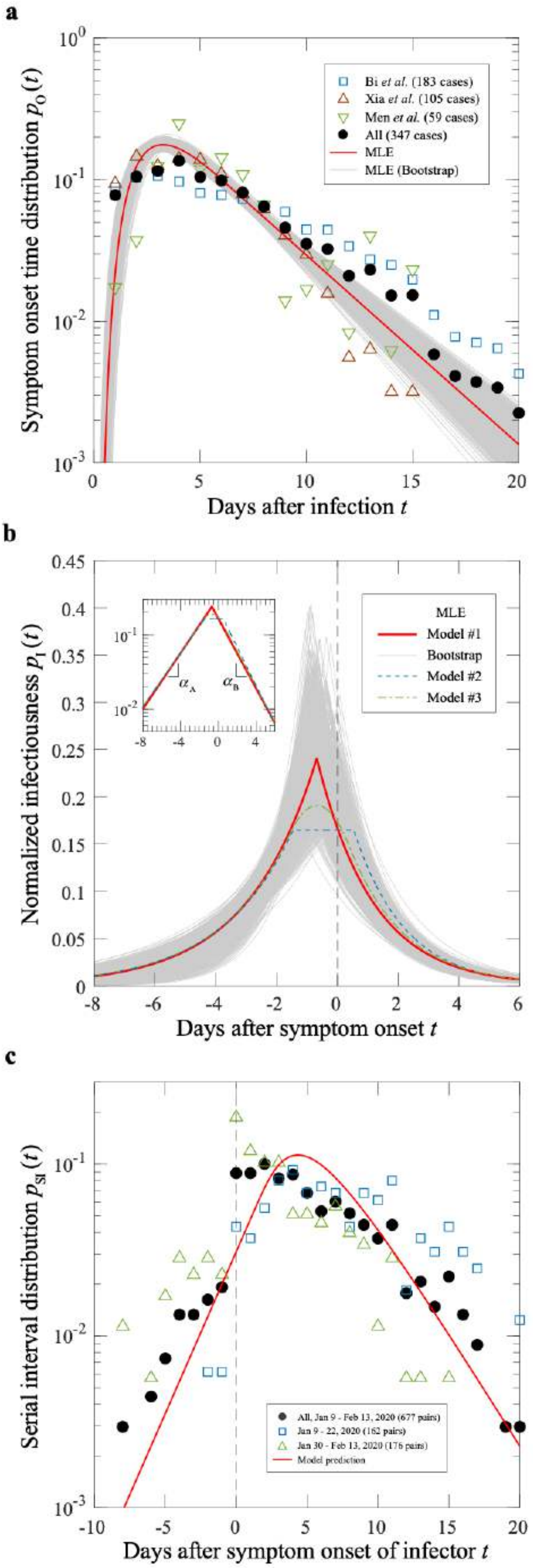
a
Symptom onset time distribution $p_O(t)$
Days after infection $t$
Bi et al. (183 cases)
Xia et al. (105 cases)
Men et al. (59 cases)
All (347 cases)
MLE
MLE (Bootstrap)
b
Normalized infectiousness $p_I(t)$
Days after symptom onset $t$
MLE
Model #1
Bootstrap
Model #2
Model #3
$\alpha_A$
$\alpha_B$
c
Serial interval distribution $p_{SI}(t)$
Days after symptom onset of infector $t$
All, Jan 9 - Feb 13, 2020 (677 pairs)
Jan 9 - 22, 2020 (162 pairs)
Jan 30 - Feb 13, 2020 (176 pairs)
Model prediction

**Figure 2. Parameter calibration from case studies. a,** The symptom onset time distribution. Raw statistics of three reported data sets (up triangle [11], down triangle [20] and square [21]) and their union (solid circle) are shown. The red curve gives the estimated distribution under a maximum likelihood scheme. Grey thin curves are generated with bootstrapping (see Methods). **b,** The infectiousness function. The data set contains 66 transmission pairs reported in Ref. 7. Result of the maximum-likelihood estimation is given by two exponential functions meeting at $-0.68$ days (red solid line). Also shown are distributions with a constant bridge (dash line), or with a dome cap (dash-dotted line), with slightly lower likelihood values (see SI Sec. 2.2). Thin grey curves are from bootstrapping for model #1 (see Methods). **c,** Serial interval statistics outside the Hubei province in China from January 9 to February 13, 2020 [22,23]: whole period (solid circles), first two weeks (open squares), and last two weeks (open triangles). The red curve is the convolution of the two red curves shown in **a** and **b**.

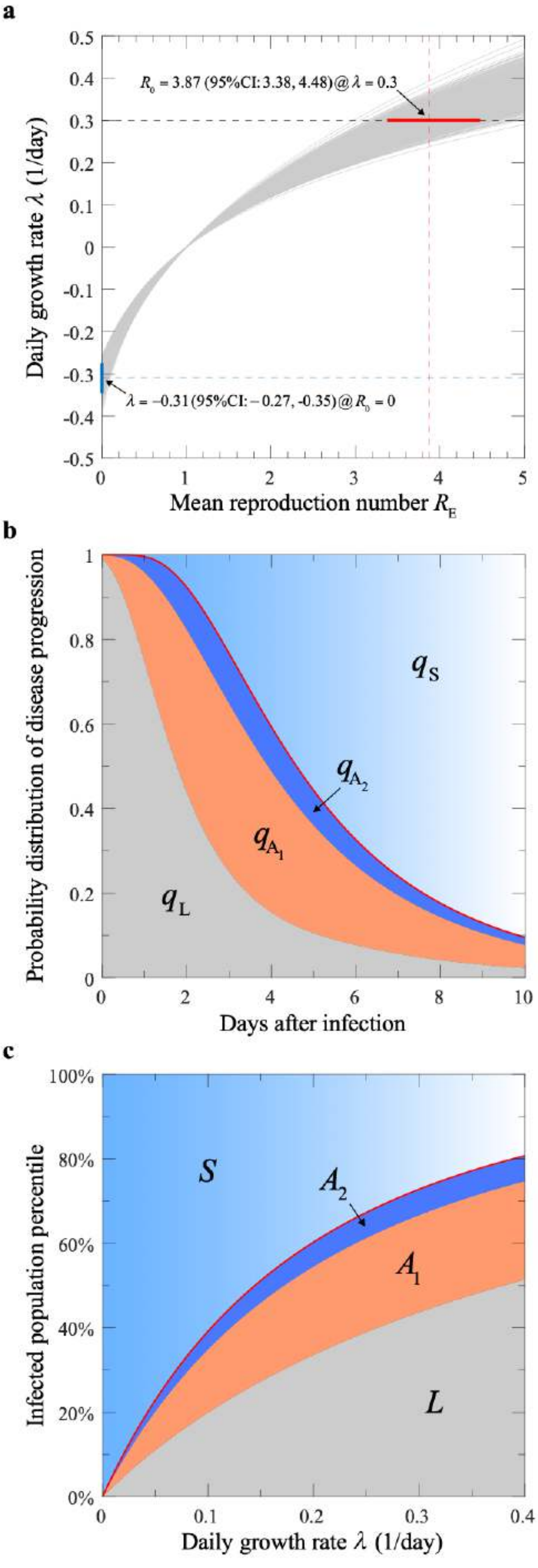
a
Daily growth rate λ (1/day)
Mean reproduction number $R_E$
$R_0$ = 3.87 (95%CI: 3.38, 4.48) @ λ = 0.3
λ = −0.31 (95%CI: − 0.27, -0.35) @ $R_0$ = 0
b
Probability distribution of disease progression
Days after infection
$q_S$
$q_{A_2}$
$q_{A_1}$
$q_L$
c
Infected population percentile
Daily growth rate λ (1/day)
S
$A_2$
$A_1$
L

**Figure 3. Basic model predictions. a,** The relationship between the growth rate $\lambda$ and mean reproduction number $R_E$. The grey lines, generated using the data shown in Fig. 2 with bootstrapping, give the range of uncertainty in the estimated function. At $\lambda = 0.3$/day, $R_E = R_0 = 3.87$. **b,** Probabilities for an infected individual being in each of the four phases on day $t$ after infection. The thick red line indicates the boundary between the pre-symptomatic and symptomatic phases. **c,** Percentage of the infected population in each of the phases when the epidemic is growing at a rate $\lambda$. The thick red curve indicates the boundary between the pre-symptomatic and symptomatic population.

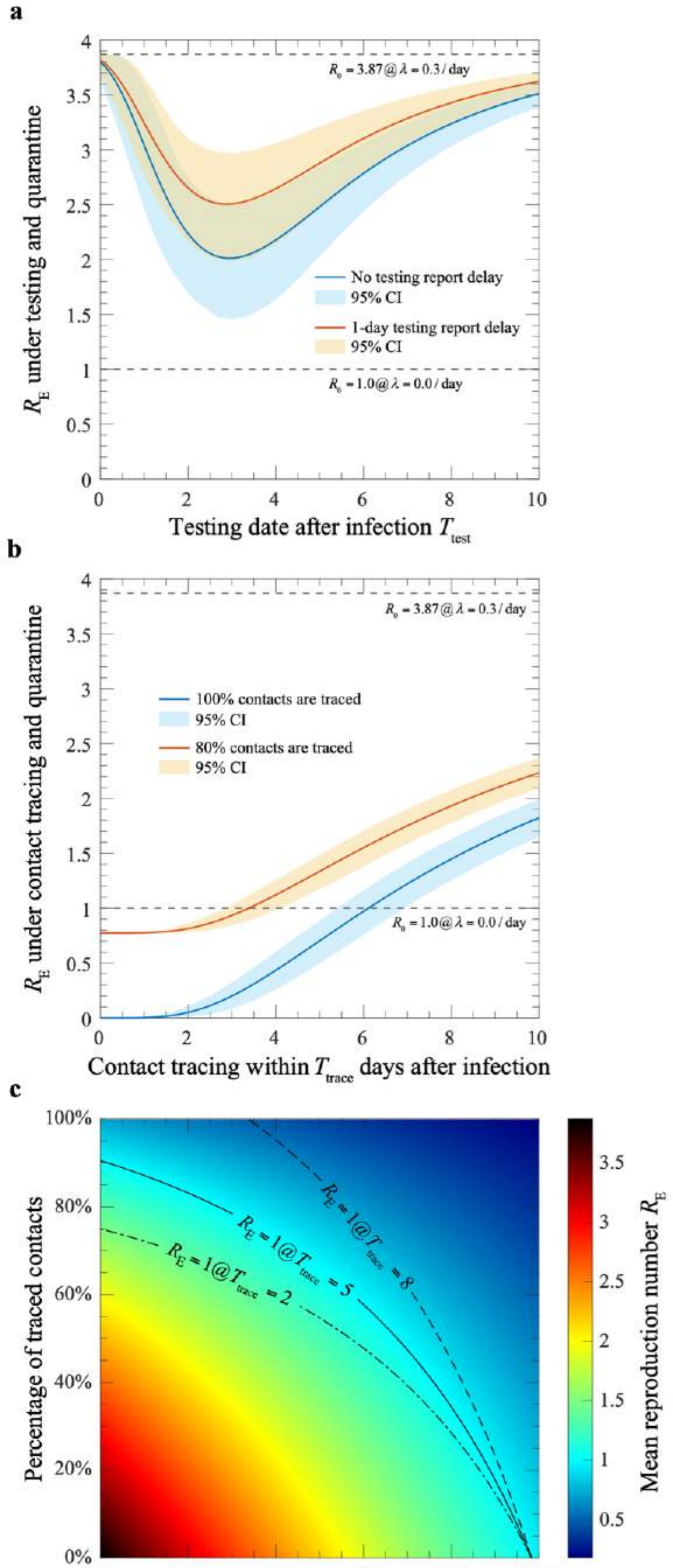
a
$R_0$ = 3.87@ λ = 0.3 / day
No testing report delay
95% CI
1-day testing report delay
95% CI
$R_0$ = 1.0@ λ = 0.0 / day
$R_E$ under testing and quarantine
Testing date after infection $T_{test}$
b
$R_0$ = 3.87@ λ = 0.3 / day
100% contacts are traced
95% CI
80% contacts are traced
95% CI
$R_0$ = 1.0@ λ = 0.0 / day
$R_E$ under contact tracing and quarantine
Contact tracing within $T_{trace}$ days after infection
c
$R_E = 1@T_{trace} = 8$
$R_E = 1@T_{trace} = 5$
$R_E = 1@T_{trace} = 2$
Percentage of traced contacts
Percentage of mask-wearing population
Mean reproduction number $R_E$

**Figure 4. Reduction of the mean reproduction number upon intervention. a,** Testing. Results are given for testing with 0 or 1 day reporting delay (blue and red curves), respectively. **b,** Contact tracing and isolation. Results are shown for 100% (blue) and 80% (red) success rates, respectively. **c,** Mask-wearing in combination with contact tracing. The heatmap gives the reduced $R_E$ when contact tracing is implemented within 5 days after infection, assuming a basal value of 3.87. The solid black line marks the percentages required to reduce $R_E$ to 1. The dash-dotted line and the dashed line map out the percentages required to flatten the epidemic growth when the time frame for contact tracing is reduced to 2 days or relaxed to 8 days, respectively. Shaded areas in **a.** and **b.** give the 95% CI of the estimated quantities.

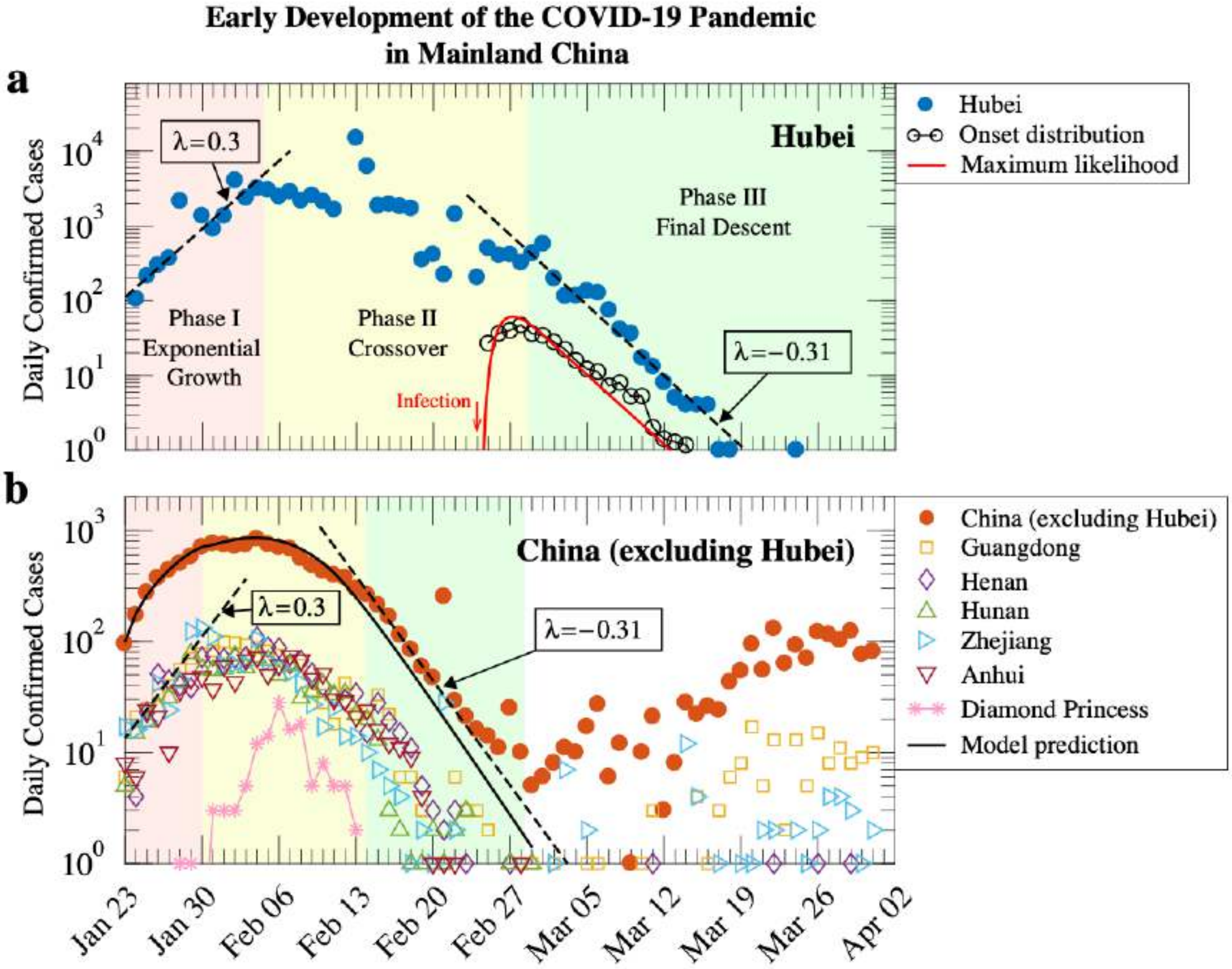

**Figure 5. Growth and containment of the COVID-19 pandemic in mainland China.** Daily confirmed cases in Hubei and other provinces since the Wuhan lockdown on January 23, 2020. **a,** Hubei province. The three phases of the epidemic development are marked in color: exponential growth (red), crossover (yellow), and descent phase (green). Early exponential growth reached a rate $\lambda$ at approximately 0.3/day (left dashed line). Growth slowed and entered the crossover phase in the middle of the second week, and reached the third phase nearly four weeks later. The final descent that began in the beginning of March is characterised by $\lambda = -0.31$/day (right dashed line). The incubation period distribution is shown in open circles (reported data of 347 cases[11,21,22]) and red line (maximum likelihood estimation) to compare with the exponential decay. Start of the incubation period is indicated by the red arrow. **b,** Other provinces in China. The epidemic development in the main affected provinces followed similar pandemic patterns. Also shown is the model prediction of the daily confirmed cases (solid line), with details given in SI Sec. 4.6. Newly confirmed cases from March onward (white region) are largely imported. Data for the Diamond Princess cruise ship[13] is included for comparison (asterisks).

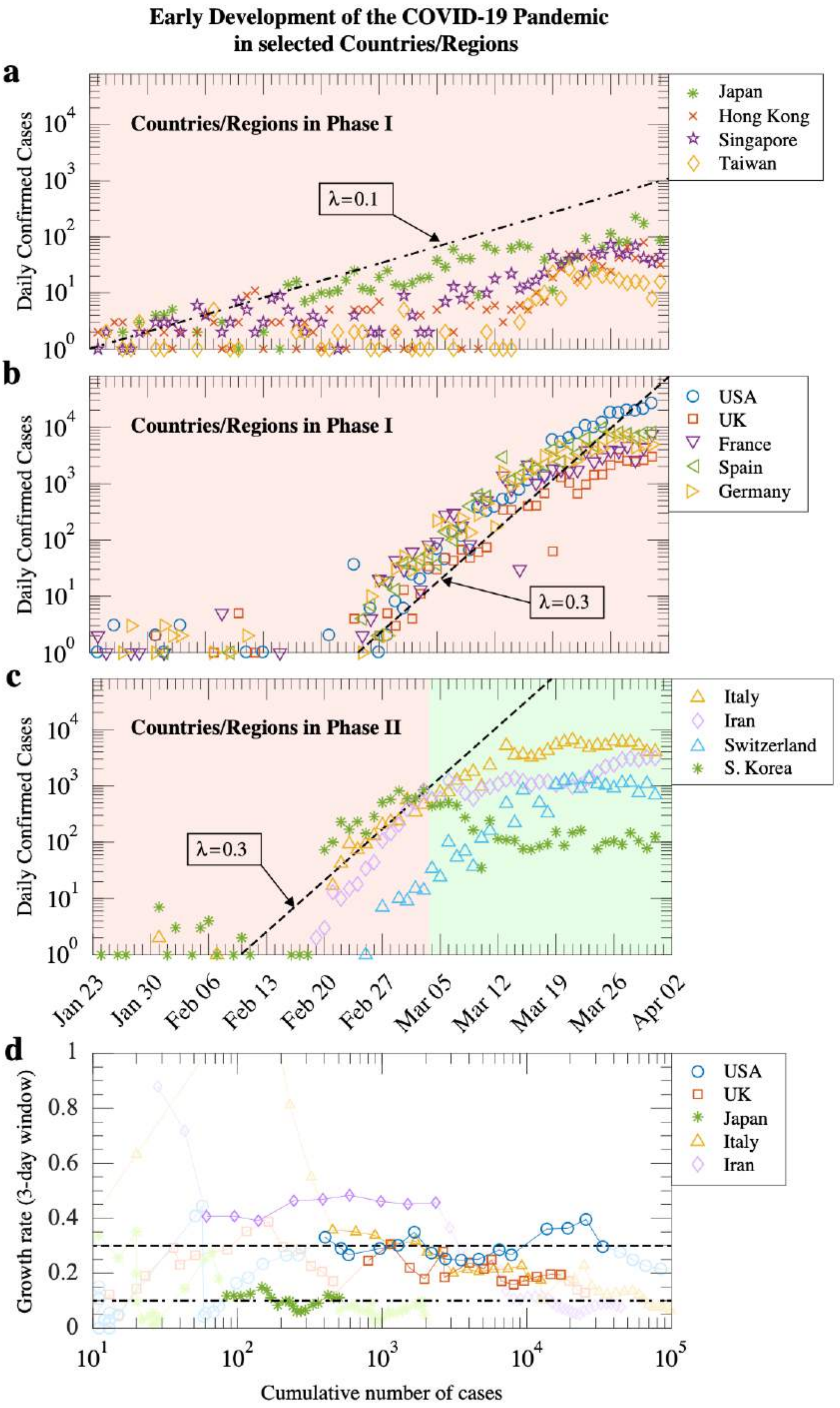

**Figure 6. First wave of the COVID-19 pandemic in selected countries and regions. a-c,** The number of daily confirmed cases from late January till end of March 2020. Countries/regions in **a** were successful in keeping transmission at a low level while those in **b** experienced exponential growth of local cases. Countries/regions in **c** have entered or been in the middle of phase II. Italy, South Korea, and Switzerland have reached zero or negative growth in daily confirmed cases, while data from Iran indicates a slowing down of the exponential growth. **d,** The estimated epidemic growth rate $\lambda(t)$ against the cumulative number of confirmed cases $N(t)$ in five representative countries. Dashed and dashed-dotted lines indicate the exponential growth rates of 0.3/day and 0.1/day, respectively.

Supplementary Information

# Calibrated Intervention and Containment of the COVID-19 Pandemic

Liang Tian, Xuefei Li, Fei Qi, Qian-Yuan Tang, Viola Tang, Jiang Liu, Zhiyuan Li, Xingye Cheng, Xuanxuan Li, Yingchen Shi, Haiguang Liu, Lei-Han Tang

In this Supplementary Information, a stochastic model of COVID-19 transmission in a homogeneous population is presented and analysed. Disease progression of an individual is compartmentalised into latent (L), pre-symptomatic infectious (A) and symptomatic (S) phases. To better accommodate the transmission characteristics of COVID-19 at the population level, the A phase is further split into two sub-periods, $A_1$ before the infectiousness peak with a variable duration, and $A_2$ after the infectiousness peak with a fixed duration. With regard to the transmission capacity, the $A_2$ phase and the S phase are treated as a single stretch of declining infectiousness. For a sufficiently large population, we derive integro-differential equations governing the size of infected subpopulations, and present analytic and numerical solutions with and without intervention.

## Contents

# 1 The Transmission Model

## 1.1 The governing equation

The basic structure of our model follows Fig. 1 in the Main Text, with model parameters defined in Fig. S1. The disease progression parameters $\alpha_{\mathrm{A}}$ and $\alpha_{\mathrm{L}}(t_{\mathrm{L}})$ are taken to be universal, while the transmission rates $\beta_{\mathrm{A}}$ and $\beta_{\mathrm{B}}(t_{\mathrm{B}})$ may vary significantly from community to community.

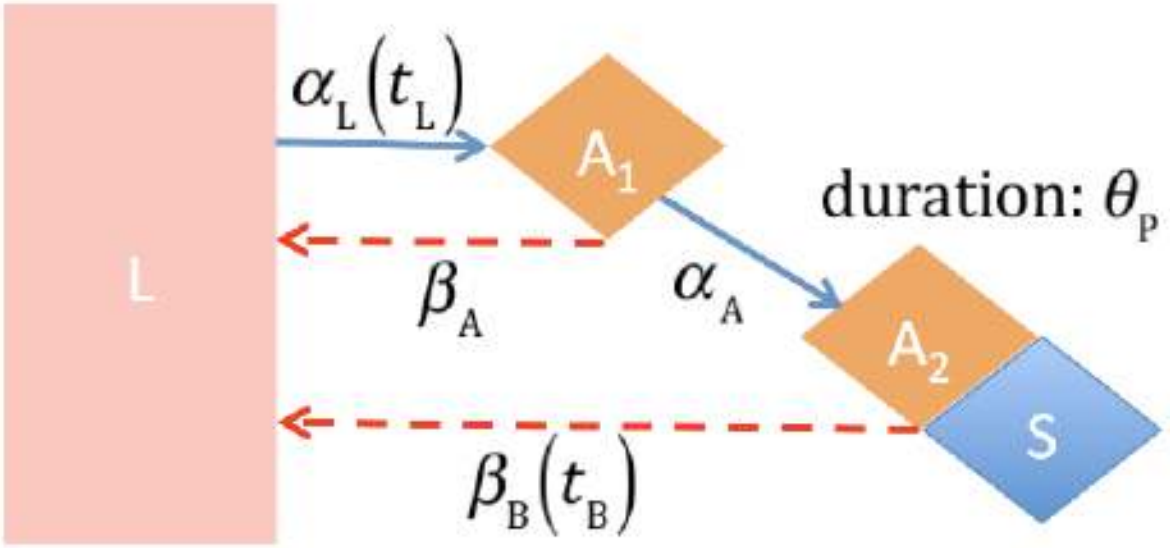

**Figure S1: A stochastic model for disease progression and transmission.** Disease progression of an infected individual is assumed to be described by a renewal process following the sequence of latent (L), pre-symptomatic infectious ($\mathrm{A}_1$ and $\mathrm{A}_2$) and symptomatic (S) phases. The transition rate from L to $\mathrm{A}_1$ is given by $\alpha_{\mathrm{L}}(t_{\mathrm{L}})$ which depends on the dwell time $t_{\mathrm{L}}$ in the latent phase. The transition from $\mathrm{A}_1$ to $\mathrm{A}_2$, which coincides with peak infectiousness, is Poisson at a constant rate $\alpha_{\mathrm{A}}$. The $\mathrm{A}_2$ phase has a fixed duration $\theta_{\mathrm{P}}$, after which the patient enters S phase. All three phases $\mathrm{A}_1$, $\mathrm{A}_2$ and S are infectious, with transmission rates to reproduce secondary cases given by $\beta_{\mathrm{A}}$ ($\mathrm{A}_1$) and $\beta_{\mathrm{B}}(t_{\mathrm{B}})$ ($\mathrm{A}_2$ and S combined), respectively. The latter is a function of $t_{\mathrm{B}}$, the number of days since the start of the $\mathrm{A}_2$ phase.

We now consider groups of infected individuals in L, $\mathrm{A}_1$, $\mathrm{A}_2$ and S in a large population, using italic symbols to denote their size. To have a complete profile of each group, one in general needs to introduce an "age counter" to specify the elapsed time since entering the disease phase. However, our assumption of the constancy of the parameters $\alpha_A$ and $\beta_A$ makes it unnecessary to do so for infected individuals in the $\mathrm{A}_1$ phase. This property greatly simplifies the description of the time evolution of the infected population as we explain below.

The generation rate of L at time $t$ is given by,

$$J_{\mathrm{L}}(t) = \beta_{\mathrm{A}} A_1(t) + \int_{-\infty}^{t} \beta_{\mathrm{B}}(t-t_2)\, \alpha_A A_1(t_2)\, dt_2, \tag{S1}$$

with $\alpha_A A_1(t_2)$ being the rate to exit from $\mathrm{A}_1$ and into $\mathrm{A}_2$. These "newly infected" eventually make their way to $\mathrm{A}_1$. The flux to $\mathrm{A}_1$ due to a group infected at a time $t_1 < t$ is given by,

$$dJ_{\mathrm{A}_1}(t) = \alpha_{\mathrm{L}}(t-t_1)\, q_{\mathrm{L}}(t-t_1)\, J_{\mathrm{L}}(t_1)\, dt_1,$$

where

$$q_{\mathrm{L}}(t) = e^{-\int_0^t \alpha_{\mathrm{L}}(t_1) dt_1}$$

is the probability that an individual infected at $t = 0$ remains in the latent phase L. Adding up contributions from all such groups, we obtain

$$J_{\mathrm{A}_1}(t) = \int_{-\infty}^{t} \alpha_{\mathrm{L}}(t-t_1)\, q_{\mathrm{L}}(t-t_1)\, J_{\mathrm{L}}(t_1)\, dt_1.$$

Making use of Eq. (S1), we then have,

$$\begin{aligned} J_{\mathrm{A}_1}(t) &= \int_{-\infty}^{t} \beta_{\mathrm{A}} \alpha_{\mathrm{L}}\left(t-t_1\right) q_{\mathrm{L}}\left(t-t_1\right) A_1\left(t_1\right) dt_1 \\ &\quad + \int_{-\infty}^{t} \alpha_{\mathrm{L}}\left(t-t_1\right) q_{\mathrm{L}}\left(t-t_1\right) dt_1 \int_{-\infty}^{t_1} \beta_{\mathrm{B}}\left(t_1-t_2\right) \alpha_{\mathrm{A}} A_1\left(t_2\right) dt_2 \\ &= \int_{-\infty}^{t} K\left(t-t_1\right) A_1\left(t_1\right) dt_1. \end{aligned} \tag{S2}$$

Here the kernel function is given by,

$$K(t) = \beta_{\mathrm{A}} \alpha_{\mathrm{L}}(t) q_{\mathrm{L}}(t) + \alpha_{\mathrm{A}} \int_0^t \alpha_{\mathrm{L}}\left(t-t_2\right) q_{\mathrm{L}}\left(t-t_2\right) \beta_{\mathrm{B}}\left(t_2\right) dt_2. \tag{S3}$$

The final equation for $A_1$ takes the form,

$$\dot{A}_1 = -\alpha_{\mathrm{A}} A_1 + \int_{-\infty}^{t} K\left(t-t_1\right) A_1\left(t_1\right) dt_1. \tag{S4}$$

## 1.2 The kernel function and the mean reproduction rate

Equation (S4) can be alternatively formulated in terms of the mean reproduction rate of a viral carrier infected at $t=0$. In our current setting,

$$\begin{aligned} r(t) &= \beta_{\mathrm{A}} \int_0^t q_{\mathrm{L}}\left(t-t_1\right) \alpha_{\mathrm{L}}\left(t-t_1\right) e^{-\alpha_{\mathrm{A}} t_1} dt_1 \\ &\quad + \int_0^t q_{\mathrm{L}}\left(t-t_1\right) \alpha_{\mathrm{L}}\left(t-t_1\right) dt_1 \int_0^{t_1} e^{-\alpha_{\mathrm{A}}\left(t_1-t_2\right)} \beta_{\mathrm{B}}\left(t_2\right) \alpha_{\mathrm{A}} dt_2 \\ &= \int_0^t K\left(t-t_1\right) e^{-\alpha_{\mathrm{A}} t_1} dt_1, \end{aligned} \tag{S5}$$

where the last step is written by comparing with Eq. (S2). It can also be verified directly from Eq. (S3).

Given the central role of Eq. (S5) in our work, we elaborate on its derivation a bit further. The distribution of the dwell time $t_{\mathrm{L}}$ in the L phase can be written as,

$$p_{\mathrm{L}}(t) \equiv \langle \delta(t-t_{\mathrm{L}}) \rangle = \alpha_{\mathrm{L}}(t)\, q_{\mathrm{L}}(t). \tag{S6}$$

Here $\langle \cdot \rangle$ denotes average over the random process of disease progression in the population, and $\delta(x)$ is the Dirac delta-function. Equation (S3) for the kernel function $K(t)$ contains two terms attributed to the transmission by a viral carrier in the $\mathrm{A}_1$ phase at $t=0$. The first term gives the rate of transmission at $t=0$, with the infected going through an incubation period of length $t$ and entering the infectious phase $\mathrm{A}_1$ at time $t$. The second term gives the rate of entering the $\mathrm{A}_2$ phase at $t=0$ and subsequent transmission at $t_2$, followed by an incubation period of $t-t_2$ for the infected. Consequently, we may write, with the help of Eq. (S6),

$$K(t) = \beta_{\mathrm{A}} \langle \delta(t-t_{\mathrm{L}}) \rangle + \alpha_{\mathrm{A}} \int_0^t \beta_{\mathrm{B}}(t_2) \langle \delta(t-t_2-t_{\mathrm{L}}) \rangle dt_2. \tag{S7}$$

On the other hand, the mean reproduction rate $r(t)$ is the rate of transmission at $t$ by an individual infected at $t=0$. This could happen if the time $t=t_{\mathrm{L}}+t_1$ is split between $t_{\mathrm{L}}$ for the latent period L and $t_1$ for the infectious period $\mathrm{A}_1$, or $t=t_{\mathrm{L}}+t_1+t_2$ that further includes $t_2$ for the post $\mathrm{A}_1$ period. Consequently,

$$r(t) = \int_0^t e^{-\alpha_{\mathrm{A}} t_1} \langle \delta(t-t_1-t_{\mathrm{L}}) \rangle \beta_{\mathrm{A}} dt_1 + \int_0^t e^{-\alpha_{\mathrm{A}} t_1} \alpha_{\mathrm{A}} dt_1 \int_0^{t-t_1} \beta_{\mathrm{B}}(t_2) \langle \delta(t-t_1-t_2-t_{\mathrm{L}}) \rangle dt_2. \tag{S8}$$

Comparing the integrand of (S8) over $t_1$ with (S7), we easily obtain (S5).

In terms of Laplace transforms defined by $\tilde{f}(\lambda) = \int_0^\infty f(t)e^{-\lambda t}dt$, the convolution integral in Eq. (S5) takes the form,

$$\tilde{r}(\lambda) = \frac{\tilde{K}(\lambda)}{\lambda + \alpha_{\mathrm{A}}}. \tag{S9}$$

Multiplying both sides of the equation by $\lambda + \alpha_{\mathrm{A}}$ and carry out the inverse Laplace transform, we have,

$$K(t) = \alpha_{\mathrm{A}} r(t) + \frac{dr(t)}{dt}. \tag{S10}$$

Note that Eq. (S5) can be considered as the solution to Eq. (S10), with $K(t)$ being the source term.

### 1.3 The symptom onset time distribution and disease progression

The symptom onset separates the $\mathrm{A}_2$ phase from the S phase. In our model, the $\mathrm{A}_2$ phase is assigned a fixed duration $\theta_{\mathrm{P}}$, the time lag between the peak infectiousness and the symptom onset. Consequently, we may express the symptom onset time as,

$$t_{\mathrm{O}} = t_{\mathrm{L}} + t_1 + \theta_{\mathrm{P}},$$

where $t_{\mathrm{L}}$ and $t_1$ are the times an infected person spent in the latent and $\mathrm{A}_1$ phases, respectively. Since the statistics of $t_1$ follows a Markov process with a time constant $1/\alpha_A$ in our model, one may infer the statistics of $t_{\mathrm{L}}$ from the statistics of $t_{\mathrm{O}}$. This then allows us to determine the kernel function $K(t)$ and other quantities of interest in terms of the symptom onset time distribution $p_{\mathrm{O}}(t)$, which can be estimated from clinical data (see the section below). The function $p_{\mathrm{O}}(t)$ is also known as the incubation period distribution. We use the two terms interchangeably both in SI and in the Main Text.

In Table S1 we give the probabilities for an individual infected at $t = 0$ to be in one of the four phases. Expressions in the last two rows are quite obvious. The transition probability current from $\mathrm{A}_1$ to $\mathrm{A}_2$ at time $t$ is given by $\alpha_{\mathrm{A}} q_{\mathrm{A}_1}(t)$, which is the same as the probability density for symptom onset at a later time $t + \theta_{\mathrm{P}}$. This gives the expression for $q_{\mathrm{A}_1}(t)$ in Table S1. The expression for $q_{\mathrm{L}}(t)$ is obtained simply by requiring that the sum of the probabilities is equal to one.

**Table S1: Probabilities for an individual to be in each disease phase at time $t$ (infection occurs at $t = 0$).**

| Phase | Probability | Expression |
|---|---|---|
| Latent L | $q_{\mathrm{L}}(t)$ | $1-\alpha_{\mathrm{A}}^{-1} p_{\mathrm{O}}(t+\theta_{\mathrm{P}}) - \int_0^{t+\theta_{\mathrm{P}}} p_{\mathrm{O}}(t_1)\, dt_1$ |
| Pre-symptomatic Infectious $\mathrm{A}_1$ | $q_{\mathrm{A}_1}(t)$ | $\alpha_{\mathrm{A}}^{-1} p_{\mathrm{O}}(t+\theta_{\mathrm{P}})$ |
| Pre-symptomatic Infectious $\mathrm{A}_2$ | $q_{\mathrm{A}_2}(t)$ | $\int_t^{t+\theta_{\mathrm{P}}} p_{\mathrm{O}}(t_1)\, dt_1$ |
| Symptomatic S | $q_{\mathrm{S}}(t)$ | $\int_0^t p_{\mathrm{O}}(t_1)\, dt_1$ |

From the expression for $q_{\mathrm{L}}(t)$ in Table S1, we obtain its Laplace transform,

$$\tilde{q}_{\mathrm{L}}(\lambda) = \frac{1}{\lambda} - \left(\frac{1}{\alpha_{\mathrm{A}}} + \frac{1}{\lambda}\right) e^{\lambda\theta_{\mathrm{P}}} \tilde{p}_{\mathrm{O}}(\lambda), \tag{S11}$$

where we have set $p_{\mathrm{O}}(t) = 0$ for $t < \theta_{\mathrm{P}}$. The Laplace transform of Eq. (S3) takes the form,

$$\tilde{K}(\lambda) \equiv \int_0^\infty K(t)e^{-\lambda t}dt = \left[\beta_{\mathrm{A}} + \alpha_{\mathrm{A}}\tilde{\beta}_{\mathrm{B}}(\lambda)\right]\left[1 - \lambda\tilde{q}_{\mathrm{L}}(\lambda)\right].$$

With the help of Eq. (S11), we obtain

$$\tilde{K}(\lambda) = \left[\beta_{\rm A} + \alpha_{\rm A}\tilde{\beta}_{\rm B}(\lambda)\right] \left(1 + \frac{\lambda}{\alpha_{\rm A}}\right) e^{\lambda\theta_{\rm P}} \tilde{p}_{\rm O}(\lambda). \tag{S12}$$

### 1.4 Exponential growth/decay

The self-sustained growth rate $\lambda$ of an epidemic can be obtained by seeking a solution $A(t) = e^{\lambda t}$ to Eq. (S4). Simple algebra gives

$$\lambda = -\alpha_{\rm A} + \tilde{K}(\lambda). \tag{S13}$$

Combining Eq. (S12) with Eq. (S13), we obtain,

$$\left[\beta_{\rm A} + \alpha_{\rm A}\tilde{\beta}_{\rm B}(\lambda)\right] \left(1 + \frac{\lambda}{\alpha_{\rm A}}\right) e^{\lambda\theta_{\rm P}} \tilde{p}_{\rm O}(\lambda) = \lambda + \alpha_{\rm A}$$

or

$$\left[\beta_{\rm A} + \alpha_{\rm A}\tilde{\beta}_{\rm B}(\lambda)\right] e^{\lambda\theta_{\rm P}} \tilde{p}_{\rm O}(\lambda) = \alpha_{\rm A}. \tag{S14}$$

In the epidemiological literature, it is customary to express the growth rate $\lambda$ in terms of the mean reproduction number

$$R_{\rm E} = \int_0^\infty r(t)dt \equiv \tilde{r}(0) = \frac{\beta_{\rm A}}{\alpha_{\rm A}} + \tilde{\beta}_{\rm B}(0), \tag{S15}$$

where we have used Eqs. (S9) and (S12). The first term on the right-hand-side gives the contribution to $R_{\rm E}$ from the pre-symptomatic $\rm A_1$ phase, and the second term from the rest. Making use of Eq. (S14), we may write (S15) as,

$$R_{\rm E} = \left(\frac{e^{-\lambda\theta_{\rm P}}}{\tilde{p}_{\rm O}(\lambda)}\right)\left(\frac{\beta_{\rm A} + \alpha_{\rm A}\tilde{\beta}_{\rm B}(0)}{\beta_{\rm A} + \alpha_{\rm A}\tilde{\beta}_{\rm B}(\lambda)}\right). \tag{S16}$$

At $\lambda = 0$, $R_{\rm E} = 1$, as required, independent of the model parameters.

Equation (S16) defines the fundamental relation between the mean reproduction number of infected individuals and the growth rate of the epidemic. The distribution $p_{\rm O}(t)$ plays an essential role in this relation. One immediate result from Eq. (S16) is that $R_{\rm E} = 0$ is at the pole of $\tilde{p}_{\rm O}(\lambda)$. For an exponentially decaying $p_{\rm O}(t) \sim e^{-\lambda t}$ at large $t$, the pole is at $\lambda = -\lambda_O$, which yields the rate of decay when transmission stops completely.

Wallinga and Lipsitch [1] proposed a general equation between $R_{\rm E}$ and $\lambda$ based on the normalised "generation interval distribution",

$$g(t) = r(t)/R_{\rm E}. \tag{S17}$$

At the observed epidemic growth rate $\lambda$, each individual produces $R_{\rm E}(\lambda)$ offspring. Consequently,

$$\tilde{g}(\lambda) = \frac{\tilde{r}(\lambda)}{R_{\rm E}} = \frac{1}{R_{\rm E}},$$

known as the Lotka–Euler estimating equation. This equation is equivalent to (S16).

**Table S2: Percentage of the infected population in each of the disease phases when the epidemic grows at a rate $\lambda$.**

| Phase | Probability | Expression |
|---|---|---|
| Latent L | $Q_{\mathrm{L}}(\lambda)$ | $1-\left(1+\frac{\lambda}{\alpha_{\mathrm{A}}}\right)e^{\lambda\theta_{\mathrm{P}}}\tilde{p}_{\mathrm{O}}(\lambda)$ |
| Pre-symptomatic Infectious $\mathrm{A}_1$ | $Q_{\mathrm{A}_1}(\lambda)$ | $\frac{\lambda}{\alpha_{\mathrm{A}}}e^{\lambda\theta_{\mathrm{P}}}\tilde{p}_{\mathrm{O}}(\lambda)$ |
| Pre-symptomatic Infectious $\mathrm{A}_2$ | $Q_{\mathrm{A}_2}(\lambda)$ | $\left(e^{\lambda\theta_{\mathrm{P}}}-1\right)\tilde{p}_{\mathrm{O}}(\lambda)$ |
| Symptomatic S | $Q_{\mathrm{S}}(\lambda)$ | $\tilde{p}_{\mathrm{O}}(\lambda)$ |

## 1.5 Percentage of subpopulations during exponential growth

Let $J_{\mathrm{L}}(t) = J_{\mathrm{L}}(0)e^{\lambda t}$ be the flux of newly infected individuals. The population size in each phase can be expressed as Laplace transforms of expressions in Table S1,

$$\begin{aligned}
L(t) &= \int_{-\infty}^{t} q_{\mathrm{L}}\left(t-t_1\right)J_{\mathrm{L}}\left(t_1\right)dt_1 = \left[\frac{1}{\lambda}-\left(\frac{1}{\alpha_{\mathrm{A}}}+\frac{1}{\lambda}\right)e^{\lambda\theta_{\mathrm{P}}}\tilde{p}_{\mathrm{O}}(\lambda)\right]J_{\mathrm{L}}(t),\\
A_1(t) &= \int_{-\infty}^{t} q_{\mathrm{A}_1}\left(t-t_1\right)J_{\mathrm{L}}\left(t_1\right)dt_1 = \frac{1}{\alpha_{\mathrm{A}}}e^{\lambda\theta_{\mathrm{P}}}\tilde{p}_{\mathrm{O}}(\lambda)J_{\mathrm{L}}(t),\\
A_2(t) &= \int_{-\infty}^{t} q_{\mathrm{A}_2}\left(t-t_1\right)J_{\mathrm{L}}\left(t_1\right)dt_1 = \frac{1}{\lambda}\left(e^{\lambda\theta_{\mathrm{P}}}-1\right)\tilde{p}_{\mathrm{O}}(\lambda)J_{\mathrm{L}}(t),\\
S(t) &= \int_{-\infty}^{t} q_{\mathrm{S}}\left(t-t_1\right)J_{\mathrm{L}}\left(t_1\right)dt_1 = \frac{1}{\lambda}\tilde{p}_{\mathrm{O}}(\lambda)J_{\mathrm{L}}(t).
\end{aligned} \tag{S18}$$

For easy reference, the percentages of subpopulations are collected in Table S2.

## 2 Model Calibration against Case Studies

In this section, we present technical details on the estimation of model parameters from case studies reported in the literature. In line with our modelling framework, we first examine the statistics of the symptom onset time and perform a maximum likelihood analysis to determine a parametric representation of the data. We then determine the mean infectiousness of COVID-19 patients from the serial interval statistics collected during the early days of the epidemic in China. These analyses yield estimates for the transmission parameters introduced in the previous section.

### 2.1 Symptom onset time distribution

The incubation periods of individual patients before symptom onset were summarised in three articles: 59 cases collected by Men *et al.*[2], 105 cases collected by Xia *et al.* [3], and 181 cases collected by Bi *et al.* [5]. In total, we examined $N = 347$ cases with their incubation periods. In most cases, the infection date can only be assigned to a time interval of more than one day. Therefore, the actual incubation period falls between $\mathrm{IPl}_i$ and $\mathrm{IPu}_i$, $i = 1, ..., N$, where $\mathrm{IPl}_i$ and $\mathrm{IPu}_i$ set the lower and upper bounds for the incubation period of case $i$, respectively.

The raw statistics of the three datasets is shown in Fig. 2a of the Main Text. This is done by simply assign equal weight to the possible values within the reported window $[\mathrm{IPl}_i, \mathrm{IPu}_i]$ for each of the patients. The nominal probability distribution of the incubation period from each data set (or the aggregated one) is then obtained by taking a simple average over all the cases within the set. Although there is general consistency with regard to the overall shape of the distributions obtained, there are also significant deviations particularly outside the peak region. While part of the variations can be attributed to statistical fluctuations when the sample size is limited, it is also plausible that systematic bias exists under the equal probability assumption when dealing with uncertainties related to finite exposure windows.

Below we perform a more refined analysis, i.e., maximum likelihood estimation, of the underlying symptom onset time distribution $p_{\mathrm{O}}(\theta, t)$, with $\theta$ being the parameter set. Following the scheme proposed by Reich *et al.* [8], we consider the following likelihood function,

$$\begin{aligned} L(\theta; \mathbf{IP}) &= \prod_{i=1}^{N} L(\theta; \mathrm{IPl}_i, \mathrm{IPu}_i), \\ L_i = L(\theta; \mathrm{IPl}_i, \mathrm{IPu}_i) &= \int_{\mathrm{IPl}_i - 0.5}^{\mathrm{IPu}_i + 0.5} p_{\mathrm{O}}(\theta, t) dt. \end{aligned} \tag{S19}$$

Eyeballing the data in Fig. 2a, we hypothesise the tail of the distribution to be an exponential function. This motivates us to consider the following two hybrid functional forms for $p_{\mathrm{O}}(\theta, t)$ and perform the maximum likelihood estimation.

#### 2.1.1 Hybrid log-normal distribution with an exponential tail

The log-normal distribution is commonly used for the incubation period in the epidemiological literature. We connect it to a simple exponential decay at $t_{\mathrm{e}}$:

$$p_{\mathrm{O}}(t) = \begin{cases} A p_{\ln}(t), & t \leq t_{\mathrm{e}}, \\ A p_{\ln}(t_{\mathrm{e}}) e^{-\gamma(t - t_{\mathrm{e}})}, & t \geq t_{\mathrm{e}}. \end{cases} \tag{S20}$$

Here the log-normal probability density function (PDF) $p_{\ln}(t)$ is parametrised by $\mu$ and $\sigma$:

$$p_{\ln}(t) = \frac{1}{t\sigma\sqrt{2\pi}} \exp\left[-\frac{(\ln t - \mu)^2}{2\sigma^2}\right]. \tag{S21}$$

Normalisation condition of $p_{\mathrm{O}}(t)$ yields

$$A = \frac{2}{1 + \operatorname{erf}\left(\frac{\ln t_{\mathrm{e}} - \mu}{\sqrt{2}\sigma}\right) + \frac{2p_{\ln}(t_{\mathrm{e}})}{\gamma}} \tag{S22}$$

where $\operatorname{erf}(x)$ is the Gauss error function. Continuity of the derivatives at $t_{\mathrm{e}}$ demands

$$\gamma = \frac{p'_{\ln}(t_{\mathrm{e}})}{p_{\ln}(t_{\mathrm{e}})}. \tag{S23}$$

Therefore, we have in total three independent parameters $\theta = (t_{\mathrm{e}}, \mu, \sigma)$ in the above model. Through numerically optimising the likelihood function with the Nelder-Mead simplex algorithm, we obtained the optimal parameter sets for different $t_{\mathrm{e}}$, which is shown in Table S3.

**Table S3:** Maximum likelihood estimation of parameters in Eq. (S20). Results are from bootstrap re-sampling of the data described in the text, with 1000 realisations (95% confidence interval shown in parentheses).

| $t_{\mathrm{e}}$ | $-\ln L$ | $\gamma$ (day$^{-1}$) | $\mu$ | $\sigma$ |
|---|---|---|---|---|
| 4 | 404.782 | 0.290<br>(0.264,0.320) | 1.349<br>(0.416,0.568) | 0.493<br>(1.324,1.372) |
| 5 | 404.034 | 0.300<br>(0.269,0.333) | 1.460<br>(1.424,1.496) | 0.551<br>(0.487,0.616) |
| ***6** | 403.301 | 0.308<br>(0.273,0.346) | 1.512<br>(1.462,1.559) | 0.578<br>(0.517,0.643) |
| 7 | 403.897 | 0.312<br>(0.279,0.351) | 1.535<br>(1.483,1.587) | 0.591<br>(0.532,0.647) |
| 8 | 404.344 | 0.315<br>(0.282,0.351) | 1.543<br>(1.487,1.599) | 0.596<br>(0.540,0.647) |
| $\infty$<br>Log-normal | 405.564 | N/A | 1.537<br>(1.479,1.597) | 0.580<br>(0.535,0.623) |

The maximal likelihood is achieved at $t_{\mathrm{e}} = 6$. It should be noted that the exponent $\gamma$ is quite stable around 0.31/day. The bottom row gives results for the full log-normal distribution ($t_{\mathrm{e}} = \infty$), which has a smaller likelihood value than the hybrid one at $t_{\mathrm{e}} = 6$.

### 2.1.2 Hybrid Weibull distribution with an exponential tail

For the second class of functions, we use the Weibull distribution to connect with an exponential tail:

$$p_{\mathrm{O}}(t) = \begin{cases} Ap_{\mathrm{wb}}(t), & t \le t_{\mathrm{e}}, \\ Ap_{\mathrm{wb}}(t_{\mathrm{e}})e^{-\gamma(t-t_{\mathrm{e}})}, & t \ge t_{\mathrm{e}}. \end{cases} \tag{S24}$$

Here the Weibull PDF $p_{\mathrm{wb}}(t)$ is parametrised by $k$ (shape parameter) and $\lambda$ (scale parameter):

$$p_{\mathrm{wb}}(t) = \frac{k}{\lambda}\left(\frac{t}{\lambda}\right)^{k-1} \exp\left[-\left(\frac{t}{\lambda}\right)^{k}\right]. \tag{S25}$$

The normalisation factor is

$$A = \frac{1}{1 - \exp\left[-\left(\frac{t_{\mathrm{e}}}{\lambda}\right)^{k}\right] + \frac{p_{\mathrm{wb}}(t_{\mathrm{e}})}{\gamma}}. \tag{S26}$$

Continuity of derivatives at $t_e$ yields

$$\gamma = \frac{p'_{wb}(t_e)}{p_{wb}(t_e)} \tag{S27}$$

This model has three independent parameters $\theta = (t_e, k, \lambda)$. We numerically optimised the likelihood function through the Nelder-Mead simplex algorithm at different $t_e$, with results given in Table S4.

**Table S4:** Maximum likelihood estimation of parameters in Eq. (S24). Results are from bootstrap re-sampling of the data described in the text, with 1000 realisations (95% confidence interval shown in parentheses).

| $t_e$ | $-\ln L$ | $\gamma$ (day$^{-1}$) | $k$ | $\lambda$ (day) |
|---|---|---|---|---|
| 3 | 405.991 | 0.283 (0.258,0.311) | 4.137 (3.409,5.029) | 3.029 (0.258,0.311) |
| ***4** | 404.332 | 0.303 (0.274,0.335) | 3.118 (2.701,3.619) | 3.917 (3.872,3.958) |
| 5 | 405.557 | 0.320 (0.287,0.357) | 2.638 (2.355,2.968) | 4.627 (4.542,4.711) |
| 6 | 406.088 | 0.338 (0.302,0.379) | 2.363 (2.146,2.608) | 5.150 (5.014,5.286) |
| 7 | 406.878 | 0.357 (0.317,0.403) | 2.189 (2.010,2.386) | 5.518 (5.337,5.704) |
| $\infty$ Weibull | 410.949 | N/A | 1.828 (1.707,1.958) | 6.206 (5.885,6.554) |

The maximal likelihood is achieved at $t_e = 4$. The likelihood for $t_e = \infty$ is lower which suggests that our hybrid function is more faithful to the data than the full Weibull distribution.

The estimated value for $\gamma$ here is very similar to the result presented in Table S3. However, the optimal likelihood obtained under the hybrid Weibull distribution is quite a bit less than that of the hybrid log-normal distribution. Therefore, we adopt the log-normal distribution with an exponential tail as the best estimate for $p_O(t)$ and use it in our numerical calculations.

#### 2.1.3 Uncertainty analysis through bootstrap re-sampling

Confidence intervals given in Tables S3 and S4 are obtained following a bootstrap scheme[8]. This is done by generating 1000 re-sampled copies of the initial 347 cases dataset. The maximum likelihood estimation is then performed for each of the re-sampled copy. Results for $p_O(t)$ so estimated are shown as the gray lines in Fig. 2a in the Main Text.

### 2.2 Transmission characteristics

#### 2.2.1 Infectiousness around symptom onset

Records of disease transmission by individual patients through the course of their disease progression are scarce in the public domain. We therefore estimated the transmission parameters defined in Fig. S1 indirectly using the reported infector-infectee pairs. Such a procedure is subject to bias arising from the data collection process. For example, intra-family transmission tends to be over-represented[4, 6]. Keeping potential shortcomings of this type in mind, we present below quantification of our model using the data provided by He *et al.* [7]

The dataset contains 74 infector-infectee pairs. Among them, 66 pairs have a unique symptom onset date (Supplementary Table 1). They are used in our analysis. As illustrated in Fig. S2, each transmission pair $i$ is associated with an exposure window $\mathrm{W}_i = [\mathrm{Wl}_i, \mathrm{Wu}_i]$, where $\mathrm{Wl}_i$ and $\mathrm{Wu}_i$ are integers that together specify an exposure window around the symptom onset of the infector (see Fig. S2).

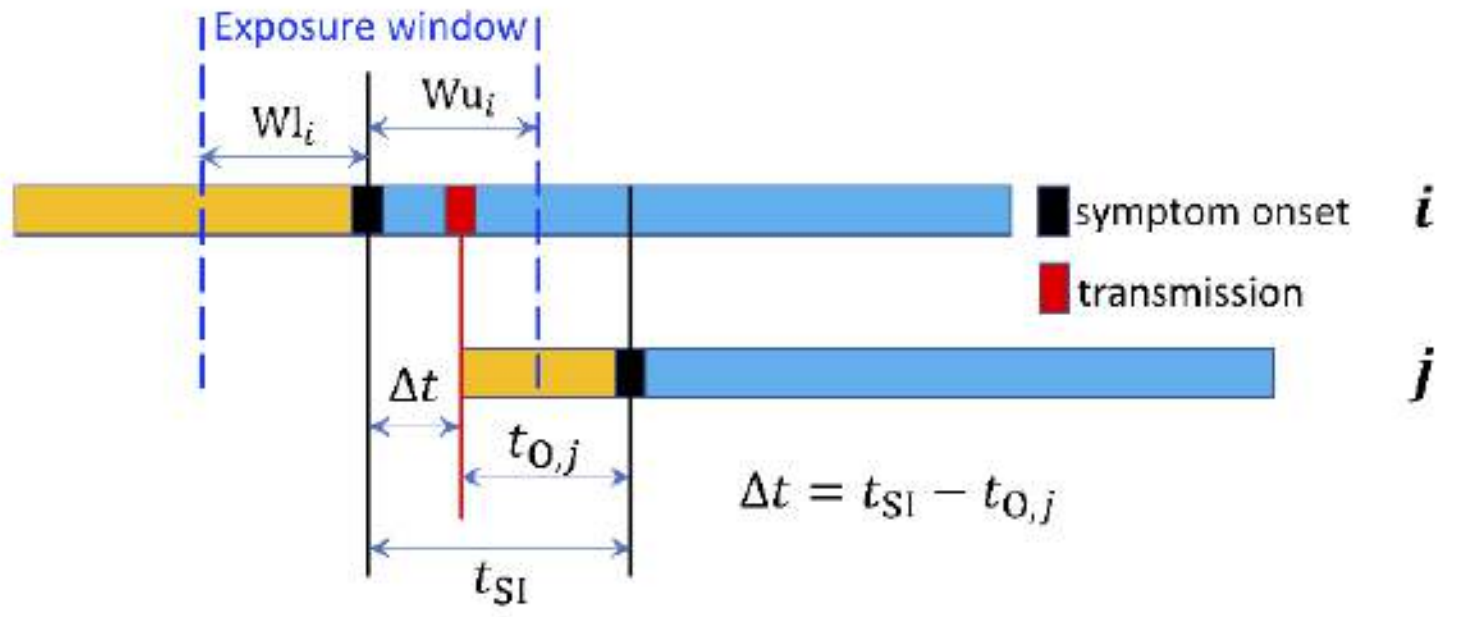

**Figure S2: Temporal events in a pairwise disease transmission.** The index patient $i$ transmits the virus to infectee $j$ within an exposure window measured from the symptom onset of patient $i$. The two symptom onsets are indicated by black bars along the time axis, while the actual transmission event is indicated by the red bar. $t_{\mathrm{SI}}$ is the serial interval, $t_{\mathrm{O},j}$ is the incubation period of infectee $j$. $\Delta t$ is the time difference between the transmission event and the symptom onset of infector $i$, which can be either negative (pre-symptomatic transmission) or positive (post-symptomatic transmission). $\mathrm{Wl}_i$ and $\mathrm{Wu}_i$ are the left and right bounds of the exposure window, w.r.t the symptom onset of infector $i$.

We quantify the statistics of disease transmission with an infectiousness function $p_{\mathrm{I}}(\theta, t)$, which gives the probability that the pairwise transmission takes place at time $t$ from the symptom onset of the infector. Then maximum likelihood estimation is applied to fit the discrete exposure window data [8]. The likelihood function is given by:

$$\begin{aligned} L(\theta; \mathbf{W}) &= \prod_{i=1}^{N} L(\theta; \mathrm{W}_i), \\ L_i = L(\theta; \mathrm{W}_i) &= \int_{\mathrm{Wl}_i - 0.5}^{\mathrm{Wu}_i + 0.5} p_{\mathrm{I}}(\theta, t) dt. \end{aligned} \tag{S28}$$

Given the limited resolution for the exposure window in integral values, we shall limit ourselves to the following parametrisations of $p_{\mathrm{I}}(\theta, t)$: a conglomerate of two exponential wings and its extensions with either a flat or a smooth cap. We consider each of them separately below.

### 2.2.2 A conglomerate of two exponential wings

We first consider the case of two exponential functions joined directly at $t_{\mathrm{P}}$:

$$p_{\mathrm{I}}(\theta, t) = \begin{cases} A e^{\alpha_{\mathrm{A}}(t - t_{\mathrm{P}})}, & t \le t_{\mathrm{P}}, \\ A e^{-\alpha_{\mathrm{B}}(t - t_{\mathrm{P}})}, & t \ge t_{\mathrm{P}}. \end{cases} \tag{S29}$$

This model has three independent parameters $\theta = (\alpha_{\mathrm{A}}, \alpha_{\mathrm{B}}, t_{\mathrm{P}})$. The normalisation factor $A$ is simply

$$A = \frac{\alpha_{\mathrm{A}} \alpha_{\mathrm{B}}}{\alpha_{\mathrm{A}} + \alpha_{\mathrm{B}}}. \tag{S30}$$

By numerically optimising the likelihood function through the Nelder-Mead simplex algorithm, we obtained the optimal parameter set shown in Table S5.

**Table S5:** Maximum likelihood estimation of the parameters in Eq. (S29) with 1000 bootstrap re-samplings (95% confidence interval shown in parentheses).

| $-\ln L$ | $\alpha_{\rm A}$ | $\alpha_{\rm B}$ | $t_{\rm P}$ |
|---|---|---|---|
| 55.551 | 0.434 (0.320,0.692) | 0.541 (0.475,0.652) | -0.677 (-1.019,-0.124) |

#### 2.2.3 Exponential wings with a flat cap

Expanding the peak of the function defined by (S29) into a flat cap of size $\epsilon$, we have,

$$p_{\rm I}(\theta,t) = \begin{cases} Ae^{\alpha_{\rm A}(t-t_{\rm A})}, & t \le t_{\rm A}, \\ A, & t_{\rm A} \le t \le t_{\rm B}, \\ Ae^{-\alpha_{\rm B}(t-t_{\rm B})}, & t \ge t_{\rm B}. \end{cases} \tag{S31}$$

Here,

$$\begin{aligned} t_{\rm A} &= t_{\rm P} - \frac{\epsilon}{2}, \\ t_{\rm B} &= t_{\rm P} + \frac{\epsilon}{2}. \end{aligned} \tag{S32}$$

The normalisation factor is simply

$$A = \frac{1}{\frac{1}{\alpha_{\rm A}} + \frac{1}{\alpha_{\rm B}} + \epsilon}. \tag{S33}$$

This model has four independent parameters $\theta = (\alpha_{\rm A}, \alpha_{\rm B}, t_{\rm P}, \epsilon)$.

Through numerical optimisation of the likelihood functions at different values of $\epsilon$, we obtained the optimal parameter sets, which are shown in Table S6.

**Table S6:** Maximum likelihood estimation of parameters in Eq. (S31).

| $\epsilon$ | $-\ln L$ | $\alpha_{\rm A}$ | $\alpha_{\rm B}$ | $t_{\rm P}$ | $t_{\rm A}$ | $t_{\rm B}$ |
|---|---|---|---|---|---|---|
| 2.0 | 56.705 | 0.417 | 0.596 | -0.460 | -1.460 | 0.540 |
| 1.0 | 55.907 | 0.428 | 0.558 | -0.630 | -1.130 | -0.130 |
| 0.5 | 55.660 | 0.434 | 0.543 | -0.697 | -0.947 | -0.447 |
| 0.2 | 55.569 | 0.434 | 0.541 | -0.690 | -0.790 | -0.590 |
| 0.1 | 55.555 | 0.434 | 0.541 | -0.683 | -0.733 | -0.633 |
| 0 | 55.551 | 0.434 | 0.541 | -0.677 | -0.677 | -0.677 |

It can be seen from the above table that the likelihood increases with decreasing cap width $\epsilon$, reaching its maximum at $\epsilon = 0$ which is treated above.

#### 2.2.4 Exponential wings with a smooth cap

We now consider the case of a smooth cap as defined by:

$$p_{\rm I}(\theta,t) = \begin{cases} A\left[1-\chi\left(t_{\rm A}-t_{\rm P}\right)^2\right] e^{\alpha_{\rm A}(t-t_{\rm A})}, & t \le t_{\rm A}, \\ A\left[1-\chi\left(t-t_{\rm P}\right)^2\right], & t_{\rm A} \le t \le t_{\rm B}, \\ A\left[1-\chi\left(t_{\rm B}-t_{\rm P}\right)^2\right] e^{-\alpha_{\rm B}(t-t_{\rm B})}, & t \ge t_{\rm B}. \end{cases} \tag{S34}$$

The shape of the quadratic function in the middle is parametrised by $\chi$, with its peak located at $t_{\mathrm{P}}$. The normalisation factor is

$$A = \frac{1}{(t_{\mathrm{B}} - t_{\mathrm{A}}) - \frac{\chi}{3}\left[(t_{\mathrm{B}} - t_{\mathrm{P}})^3 + (t_{\mathrm{P}} - t_{\mathrm{A}})^3\right] + \frac{1}{\alpha_{\mathrm{A}}}\left[1 - \chi(t_{\mathrm{A}} - t_{\mathrm{P}})^2\right] + \frac{1}{\alpha_{\mathrm{B}}}\left[1 - \chi(t_{\mathrm{B}} - t_{\mathrm{P}})^2\right]}. \tag{S35}$$

Smoothness requires continuity of derivatives at $t_{\mathrm{A}}$ and $t_{\mathrm{B}}$, respectively, which yields:

$$\begin{aligned} \alpha_{\mathrm{A}} &= \frac{2\chi(t_{\mathrm{P}} - t_{\mathrm{A}})}{1 - \chi(t_{\mathrm{P}} - t_{\mathrm{A}})^2}, \quad t_{\mathrm{A}} = t_{\mathrm{P}} + \frac{1}{\alpha_{\mathrm{A}}} - \sqrt{\frac{1}{\alpha_{\mathrm{A}}^2} + \frac{1}{\chi}}, \\ \alpha_{\mathrm{B}} &= \frac{2\chi(t_{\mathrm{B}} - t_{\mathrm{P}})}{1 - \chi(t_{\mathrm{B}} - t_{\mathrm{P}})^2}, \quad t_{\mathrm{B}} = t_{\mathrm{P}} - \frac{1}{\alpha_{\mathrm{B}}} + \sqrt{\frac{1}{\alpha_{\mathrm{B}}^2} + \frac{1}{\chi}}. \end{aligned} \tag{S36}$$

Therefore, there are four independent parameters $\theta = (\alpha_{\mathrm{A}}, \alpha_{\mathrm{B}}, t_{\mathrm{P}}, \chi)$. Note that in the limit of $\chi \to \infty$ this model reduces to the first model with two exponential functions joining at $t_{\mathrm{P}}$.

We numerically optimised the likelihood functions through the Nelder-Mead simplex algorithm at different $\chi$. Optimal parameter sets in each case are shown in Table S7.

**Table S7:** Maximum likelihood estimation of parameters in Eq. (S34).

| $\chi$ | $-\ln L$ | $\alpha_{\mathrm{A}}$ | $\alpha_{\mathrm{B}}$ | $t_{\mathrm{P}}$ | $t_{\mathrm{A}}$ | $t_{\mathrm{B}}$ |
|---|---|---|---|---|---|---|
| 0.2 | 56.018 | 0.419 | 0.551 | -0.672 | -1.556 | 0.392 |
| 0.5 | 55.662 | 0.431 | 0.541 | -0.707 | -1.104 | -0.228 |
| 1.0 | 55.584 | 0.434 | 0.540 | -0.704 | -0.912 | -0.452 |
| 5.0 | 55.552 | 0.434 | 0.541 | -0.683 | -0.726 | -0.629 |
| 10.0 | 55.551 | 0.434 | 0.541 | -0.680 | -0.702 | -0.653 |
| $\infty$ | 55.551 | 0.434 | 0.541 | -0.677 | -0.677 | -0.677 |

Note that the likelihood increases with increasing $\chi$, which indicates that the optimal estimation is achieved at vanishing cap size.

In Fig. 2b in the Main Text, we show the infectiousness curves obtained from the first model, the second model at $\epsilon = 1.0$, and the third model at $\chi = 1.0$.

#### 2.2.5 Uncertainty analysis through bootstrap re-sampling

We performed bootstrap analysis to determine uncertainties in the estimated $p_{\mathrm{I}}(t)$. This is done by generating 1000 re-sampled copies of the exposure window dataset of 66 transmission pairs. The maximum likelihood estimation for $p_{\mathrm{I}}(t)$ is then performed for each of the re-sampled copy. The $p_{\mathrm{I}}(t)$ for each of the re-sampled copied are shown as the grey lines in Main Text Fig. 2b (for the conglomerate of two exponentials). The 95% confidence intervals for the estimated parameters are shown in Table S5 and the Main Text Table I.

## 3 Simplifying Approximations

In Sec. 1 we expressed various quantities in terms of the Laplace transform of $p_{\rm O}(t)$. Here we consider simplifying approximations, which allow for more direct relations to be derived.

### 3.1 An approximate formula for the mean reproduction rate

As an explicit example, we consider disease transmission in a homogeneous population following the infectiousness curve shown in Fig. 2b. The peak infectiousness happens at $t_{\rm P} = -0.68$ days before the symptom onset, yielding $\theta_{\rm P} = 0.68$ days. The left and right wings are well fitted to exponential functions with decay rates $\alpha_{\rm A} \simeq 0.43$ day$^{-1}$ and $\alpha_{\rm B} \simeq 0.54$ day$^{-1}$, respectively. For the right wing, we may write

$$\beta_{\rm B}(t_{\rm B}) = \beta_{\rm A} \exp(-\alpha_{\rm B} t_{\rm B}), \tag{S37}$$

where $t_{\rm B}$ is measured from the beginning of the $\rm A_2$ phase, e.g., $\theta_{\rm P}$ days before the symptom onset. The mean infectiousness curve on the left also matches well with the exponential distribution of the duration of the $\rm A_1$ phase in the model introduced in Sec. 1, with a decay rate $\alpha_{\rm A}$.

From Eq. (S37), we obtain the Laplace transform,

$$\tilde{\beta}_{\rm B}(\lambda) = \int_0^\infty \beta_{\rm A} e^{-\alpha_{\rm B} t} e^{-\lambda t} dt = \frac{\beta_{\rm A}}{\lambda + \alpha_{\rm B}}. \tag{S38}$$

Substituting Eq. (S38) into (S12), we obtain,

$$\tilde{K}(\lambda) = \beta_{\rm A} \left(1 + \frac{\alpha_{\rm A}}{\lambda + \alpha_{\rm B}}\right) \left(1 + \frac{\lambda}{\alpha_{\rm A}}\right) e^{\lambda \theta_{\rm P}} \tilde{p}_{\rm O}(\lambda). \tag{S39}$$

The Laplace transform of the mean reproduction rate $r(t)$ can be obtained from $\tilde{K}(\lambda)$ using Eq. (S9). For $\lambda \ll \alpha_{\rm B}$, we may approximate

$$1 + \frac{\alpha_{\rm A}}{\lambda + \alpha_{\rm B}} \simeq \left(1 + \frac{\alpha_{\rm A}}{\alpha_{\rm B}}\right) \left(1 - \frac{\alpha_{\rm A}}{\alpha_{\rm B}} \frac{\lambda}{\alpha_{\rm A} + \alpha_{\rm B}}\right) \simeq \left(1 + \frac{\alpha_{\rm A}}{\alpha_{\rm B}}\right) e^{-\frac{\lambda \alpha_{\rm A}}{\alpha_{\rm B}(\alpha_{\rm A} + \alpha_{\rm B})}}.$$

Introducing

$$\theta_{\rm S} = \theta_{\rm P} - \frac{\alpha_{\rm A}}{\alpha_{\rm B}(\alpha_{\rm A} + \alpha_{\rm B})}, \tag{S40}$$

we may then write,

$$\tilde{r}(\lambda) \simeq \left(\frac{\beta_{\rm A}}{\alpha_{\rm A}} + \frac{\beta_{\rm A}}{\alpha_{\rm B}}\right) e^{\lambda \theta_{\rm S}} \tilde{p}_{\rm O}(\lambda). \tag{S41}$$

This yields a very simple expression for the reproduction rate,

$$r(t) \simeq R_{\rm E} p_{\rm O}(t + \theta_{\rm S}), \tag{S42}$$

with $R_{\rm E} = (\beta_{\rm A}/\alpha_{\rm A}) + (\beta_{\rm A}/\alpha_{\rm B})$. Equation (S17) then identifies $p_{\rm O}(t + \theta_{\rm S})$ with the generation time interval distribution $g(t)$.

A few remarks with regard to the re-parameterised model (S42) are in order. The parameter $R_{\rm E}$, which sets the overall scale of transmission, incorporates total contribution from both pre-symptomatic and symptomatic individuals (Fig. 1, Main Text). The shift parameter $\theta_{\rm S}$, on the other hand, depends on how fast transmission decays on the symptomatic side. As seen from Eq. (S40), a small decay rate $\alpha_{\rm B}$ may change $\theta_{\rm S}$ into the negative. Using the parameter values determined from Sec. 2, we obtain $\theta_{\rm S} \simeq -0.15$ day. In this case, the peak-shift of the infectiousness curve to the pre-symptomatic side is compensated by transmission from symptomatic patients. As reported in Ref. [9], the latter could vary over time depending on the isolation measures applied to symptomatic patients.

### 3.2 A Markov model

A number of modelling studies in the literature adopt a Markovian setup with $\alpha_{\rm L}(t) = const.$ which is a special case of our more general non-Markovian approach. Skipping the $\rm A_2$ phase by setting $\theta_{\rm P} = 0$, we have,

$$\tilde{q}_{\rm L} = \frac{1}{\alpha_{\rm L} + \lambda}, \quad \tilde{p}_{\rm O} = \frac{\alpha_{\rm L}}{\alpha_{\rm L} + \lambda}\frac{\alpha_{\rm A}}{\alpha_{\rm A} + \lambda}. \tag{S43}$$

The onset time distribution in this case is given by

$$p_{\rm O}(t) = \frac{\alpha_{\rm L}\alpha_{\rm A}}{\alpha_{\rm A} - \alpha_{\rm L}}\left[e^{-\alpha_{\rm L}t} - e^{-\alpha_{\rm A}t}\right] = \frac{1}{\tau_{\rm L} - \tau_{\rm A}}\left(e^{-t/\tau_{\rm L}} - e^{-t/\tau_{\rm A}}\right),$$

with its mean and variance given by,

$$\tau_{\rm O} = \langle t_{\rm O}\rangle = \tau_{\rm L} + \tau_{\rm A}, \quad \sigma_{\rm O}^2 = \langle t_{\rm O}^2\rangle - \langle t_{\rm O}\rangle^2 = \tau_{\rm L}^2 + \tau_{\rm A}^2,$$

and

$$\frac{\sigma_{\rm O}^2}{\tau_{\rm O}^2} = \varepsilon^2 + (1-\varepsilon)^2,$$

where $\tau_{\rm L} = \varepsilon\tau_{\rm O}$, $\tau_{\rm A} = (1-\varepsilon)\tau_{\rm O}$. In the limit $\tau_{\rm L} = \tau_{\rm A} =$ or $\varepsilon = 1/2$, $p_{\rm O}(t) = \frac{t}{\tau_{\rm L}^2}e^{-t/\tau_{\rm L}}$.

Substituting Eq. (S43) into Eq. (S12), we obtain the Laplace transform of the kernel function,

$$\tilde{K}_{\rm M}(\lambda) = \alpha_{\rm L}\frac{\beta_{\rm A} + \alpha_{\rm A}\tilde{\beta}_{\rm B}(\lambda)}{\alpha_{\rm L} + \lambda}. \tag{S44}$$

Below we consider a situation where $\beta_{\rm B}(t)$ decays to zero rapidly, i.e., the symptomatic transmission is much reduced. Expanding the numerator in (S44) up to the first order in $\lambda$, we obtain,

$$\beta_{\rm A} + \alpha_{\rm A}\tilde{\beta}_B(\lambda) \simeq \beta_{\rm eff}(1 - \tau_{\rm eff}\lambda),$$

where $\beta_{\rm eff} = \beta_{\rm A} + \alpha_{\rm A}\tilde{\beta}_{\rm B}(0)$ and $\tau_{\rm eff} = -\alpha_{\rm A}\tilde{\beta}'_{\rm B}(0)/\beta_{\rm eff}$. It follows that

$$\tilde{K}_{\rm M}(\lambda) \simeq \alpha_{\rm L,eff}\frac{\beta_{\rm eff}}{\alpha_{\rm L,eff} + \lambda}, \tag{S45}$$

where $\alpha_{L,\rm eff} = \alpha_{\rm L}/\left(1 + \alpha_{\rm L}\tau_{\rm eff}\right)$ gives the reduced effective rate to exit the latent phase L, i.e., the lifetime of the latent phase is extended. The inverse transform of the above expression yields

$$K_{\rm M}(t) \simeq \beta_{\rm eff}\alpha_{\rm L,eff}e^{-\alpha_{\rm L,eff}t}. \tag{S46}$$

Under the effective kernel function (S46), Eq. (S13) takes the form,

$$R_{\rm E} = \left(1 + \frac{\lambda}{\alpha_{\rm L,eff}}\right)\left(1 + \frac{\lambda}{\alpha_{\rm A}}\right),$$

i.e., a parabola with two nodes at $\lambda_L \simeq -\alpha_{\rm L,eff}$ and $\lambda_{\rm A} \simeq -\alpha_{\rm A}$.

## 4 Model Exploration under Intervention

In this section, we consider the effects of various intervention and containment measures aimed at a significant reduction of the mean reproduction number $R_{\mathrm{E}}$ from its nominal number of 3 or above in an unprotected population, within the framework of the model described in Sec. 1. When these measures are implemented in combination, reductions multiply. Some of these measures, such as social distancing and wearing face masks, are aiming at an overall reduction of COVID-19 transmission in the population, while others are targeting more specifically at breaking down the transmission chain. Considering the difficulty in achieving a three-fold reduction of $R_{\mathrm{E}}$ under any single measure, one is left with no option other than adopting as many of these measures as circumstances allow.

### 4.1 Reduction in the mean reproduction number upon quarantine

In the following discussion we shall assume that, once a viral carrier is identified, quarantine measures will take effect which terminate disease transmission by the individual in question. Let $t$ be the time interval between infection and identification. The mean transmission reduction per individual is given in Table S8.

**Table S8: Transmission reduction for an individual infected at $t = 0$ and quarantined at time $t$.**

| Disease Phase on the Day of Quarantine | Mean Reduction | Result |
|---|---|---|
| L & $A_1$ | $\Delta R_{\mathrm{A}}$ | $R_{\mathrm{E}}$ |
| entering $A_2$ at $t_1 < t$ | $\Delta R_{\mathrm{B}}(t - t_1)$ | $\int_{t-t_1}^{\infty} \beta_{\mathrm{B}}(t_2) dt_2$ |

### 4.2 Testing and quarantine

Testing and quarantining of infected individuals is practiced in South Korea during the early stages of the pandemic with great intensity. In the simplest scenario, a suspected individual undergoes one time test of COVID-19 infection. It takes a day or so for the test result to come back. If it is positive, the person will be quarantined and hence removed from the active infected population. Oral nucleic acid test only reports cases with a sufficiently high viral load. Therefore the test needs to be done around the time of symptom onset. However, by then the person may have already infected other people. Within our probabilistic framework, the efficiency of this procedure can be assessed as follows.

We assume that the test protocol is implemented in such a way that a given individual infected at $t = 0$ is tested at a rate $\eta_{\mathrm{testing}}(t)$, i.e., the probability that he/she is tested during the time interval $(t, t + dt)$ is $\eta_{\mathrm{testing}}(t)dt$, with $\int_0^\infty \eta_{\mathrm{testing}}(t)dt \leq 1$. We shall also assume that the test result, which is returned after $\tau_{\mathrm{d}}$ days, will be positive only when the infected has already passed the latent phase on the day of testing. With the help of the results in Table S8, we may write the reduction in the mean reproduction number $R_{\mathrm{E}}$ at the population level,

$$\Delta R_{\mathrm{testing}} = \int_0^\infty dt \eta_{\mathrm{testing}}(t) \Big( q_{\mathrm{A}_1}(t) \big[ e^{-\alpha_{\mathrm{A}} \tau_{\mathrm{d}}} R_{\mathrm{E}} + \phi(\tau_{\mathrm{d}}) \big] + \int_0^t dt_1 \alpha_{\mathrm{A}} q_{\mathrm{A}_1}(t_1) \Delta R_{\mathrm{B}}(t + \tau_{\mathrm{d}} - t_1) \Big). \tag{S47}$$

Here

$$\phi(\tau) = \int_0^\tau dt \alpha_{\mathrm{A}} e^{-\alpha_{\mathrm{A}} t} \Delta R_{\mathrm{B}}(\tau - t) = \big(1 - e^{-\alpha_{\mathrm{A}} \tau}\big) \tilde{\beta}_B(0) - \int_0^\tau dt \big(1 - e^{-\alpha_{\mathrm{A}} t}\big) \beta_{\mathrm{B}}(t).$$

In the extreme case that all infected are tested on day $T$ of their infection, we have $\eta_{\mathrm{testing}}(t) = \delta(t-T)$. Under Eqs. (S37) and (S47), the relative reduction of $R_{\mathrm{E}}$ is given by,

$$\begin{aligned}\frac{\Delta R_{\mathrm{testing}}(T)}{R_{\mathrm{E}}} =& q_{\mathrm{A}_1}(T)\left(\frac{\alpha_{\mathrm{B}}}{\alpha_{\mathrm{A}}+\alpha_{\mathrm{B}}}e^{-\alpha_{\mathrm{A}}\tau_{\mathrm{d}}} + \frac{\alpha_{\mathrm{A}}}{\alpha_{\mathrm{A}}+\alpha_{\mathrm{B}}}e^{-\alpha_{\mathrm{B}}\tau_{\mathrm{d}}} + \frac{\alpha_{\mathrm{A}}\alpha_{\mathrm{B}}}{(\alpha_{\mathrm{A}}+\alpha_{\mathrm{B}})^2}\left(1-e^{-(\alpha_{\mathrm{A}}+\alpha_{\mathrm{B}})\tau_{\mathrm{d}}}\right)\right) \\ &+ \frac{\alpha_{\mathrm{A}}^2}{\alpha_{\mathrm{A}}+\alpha_{\mathrm{B}}}e^{-\alpha_{\mathrm{B}}\tau_{\mathrm{d}}}\int_0^T q_{\mathrm{A}_1}(T-t)e^{-\alpha_{\mathrm{B}}t}dt.\end{aligned} \tag{S48}$$

According to the expression in Table S1, $q_{\mathrm{A}_1}(T)$ is peaked one day before the peak of the symptom onset time distribution. Consequently, the maximum reduction is also achieved when $T$ is chosen to be around that day. For smaller $T$, the chance of returning a positive test result is low, while for larger $T$, the patient in question is likely to have already infected others by the time of the test.

### 4.3 Contact tracing

In contact tracing, a high percentage of close contacts of a newly confirmed viral carrier (primary case) are identified and quarantined soon after the contact took place, without testing. This procedure is more effective when performed sufficiently close to the contact date, as it also covers secondary cases that are still in the latent phase of their disease progression. For the secondary cases, their $r(t)$ is truncated on the day they are located, while transmission to tertiary cases could happen before that time. At an overall success rate $q_{\mathrm{c}}$, we assume infectees are traced down within a time window $T$ since infection. The reduction of $R_E$ is then given by,

$$\Delta R_{\mathrm{E}} = q_{\mathrm{c}}\int_0^T \frac{dt}{T}\int_t^\infty r(t_1)dt_1 = q_{\mathrm{c}}R_{\mathrm{E}}\Big[1-\frac{1}{T}\int_0^{T+\theta_{\mathrm{S}}} q_{\mathrm{S}}(t)dt\Big], \tag{S49}$$

where we have used Eq. (S42). Expression for $q_{\mathrm{S}}(t)$ is given in Table S1.

### 4.4 Mask wearing

Considering the dual-effects of mask-wearing in reducing both virus inhalation by susceptible individuals and exhalation by infectious individuals (including pre-symptomatic and asymptomatic groups), we calculated the reduction of transmission rate $\beta$ against mask efficacy and the percentage of the population wearing masks under a simplifying approximation.

We assumed that there are COVID-19 positive (P) and negative (N) individuals. $p_{\mathrm{m}}$ is the percentage of the population that wear masks: $p_{\mathrm{mP}}$ for positive individuals and $p_{\mathrm{mN}}$ for negative individuals. Here, $e$ denotes the efficacy of masks measured by the percentage of virus trapped by the mask: from inhalation ($e_{\mathrm{in}}$ , important for COVID-19 positive individuals) and exhalation ($e_{\mathrm{ex}}$, important for COVID-19 negative individuals). In the absence of masks, the rate of transmission by a positive individual contacting a negative individual is $\beta$.

**Table S9: Four types of encounters between a positive and a negative individual and reduction of the transmission rate $\beta$.**

| Prob. of contact type | P wearing mask | N wearing mask | Chance of transmission |
|---|---|---|---|
| $p_{\mathrm{mP}}\cdot p_{\mathrm{mN}}$ | Yes | Yes | $\beta(1-e_{\mathrm{ex}})(1-e_{\mathrm{in}})$ |
| $p_{\mathrm{mP}}(1-p_{\mathrm{mN}})$ | Yes | No | $\beta(1-e_{\mathrm{ex}})$ |
| $(1-p_{\mathrm{mP}})p_{\mathrm{mN}}$ | No | Yes | $\beta(1-e_{\mathrm{in}})$ |
| $(1-p_{\mathrm{mP}})(1-p_{\mathrm{mN}})$ | No | No | $\beta$ |

Therefore, the averaged chance of transmission is:

$$\beta_{\mathrm{mask}} = \beta(1-e_{\mathrm{in}}\cdot p_{\mathrm{mN}})\cdot(1-e_{\mathrm{ex}}\cdot p_{\mathrm{mP}}).$$

Under the totally symmetric assumption $e_{\rm ex} = e_{\rm in} = e$, $p_{\rm mN} = p_{\rm mP} = p_{\rm m}$, the result becomes:

$$\beta_{\rm mask} = \beta(1 - e \cdot p_{\rm m})^2. \tag{S50}$$

Note that this equation can be applied separately to pre-symptomatic and symptomatic transmission, with their own $p_{\rm m}$. In the numerical examples presented below and in the Main Text, we used the same $p_{\rm m}$ for both types of transmission. In the more general case, the reduction of $\beta$ can be calculated when contributions of the two types to the total are known.

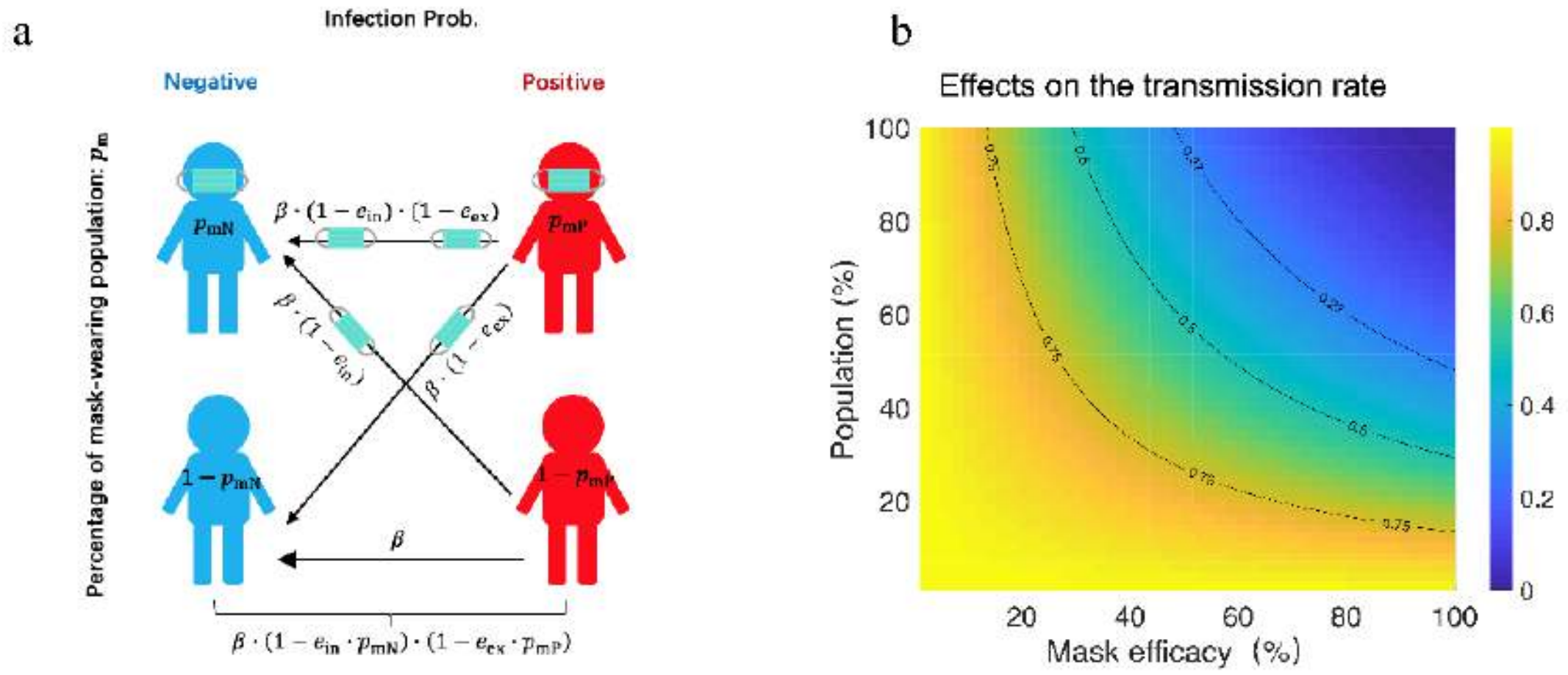

**Figure S3: Estimation on the impact of mask-wearing on the transmission rate.** a. Schematic representation on the dual-effects of masks in reducing $\beta$. b. Relationship between the factor multiplying $\beta$ (heatmap color) with the mask efficacy ($x$-axis) and the fraction of the population wearing masks ($y$-axis). Black lines show the contours for reducing $\beta$ to 0.75, 0.5, 0.27 of its original value.

The above discussion shows, in a semi-quantitative way, that mask-wearing brings benefits to one-self if not infected but more importantly to others. Even with masks at a moderate efficacy of 50%, reduction of the transmission rate can be substantial when practiced by the whole population.

Previous research on influenza suggested that surgical mask reduces 70% of the viral aerosol shedding [10]. Also, WHO suggested that respiratory droplets ($> 5 \sim 10$ $\mu$m in diameter) and contacts are the primary routes for COVID-19 to transmit between people [11]. Surgical mask reduces more than 90% of droplets in this size range [12]. However, general public may not be able to fully comply with the usage guidance of surgical masks. Therefore, in generating results in the Main Text, we take a simpler assumption that the efficacy of surgical masks is at 50%.

Regarding potential concerns on the effectiveness of mask-wearing in reducing the epidemics, we provided additional references and descriptions in the following.

In the laboratory setting, there is evidence that masks are able to filter in the relevant droplet size range for COVID-19, as well as efficacy in blocking droplets and particles from the wearer in a range higher or near the efficacy of 50% [13, 14]. For seasonal coronaviruses, surgical masks for source control were effective at blocking coronavirus droplets of all sizes for every subject [15]. Personal protection is more challenging than source control, since the inhaling particles are smaller. According to World Health Organisation's "Advice on the use of masks in the context of COVID-19" [16], the penetration for surgical masks is 50%-60% , which is the range we used in our simplified model.

There are already several experimental measurements on the efficacy of different types of masks against coronavirus, both as source control [15, 17, 18] and personal protection equipment [16, 19, 20]. In summary, there is laboratory-based evidence that surgical or N95 masks have satisfying filtration capacity in the relevant droplet size range of coronavirus. The experimental reports are included in the references.

Several recent modelling works focusing on the effect of population-wide mask-wearing converge to similar conclusions that masks of intermediate filtering efficacy exhibit aggregate effect at the population level. For example, in the work of Stutt *et al.* [21], they found that with a policy that all individuals must wear a mask all of the time, a median effective COVID-19 $R_{\mathrm{E}}$ of below 1 could be reached, even with mask effectiveness of 50% (for $R_{\mathrm{E}} = 2.2$) or mask effectiveness of 75% (for $R_{\mathrm{E}} = 4$). Similarly, models from Kai *et al.* [22] estimated that 80-90% masking would eventually eliminate the disease. Work from Fisman *et al.* [23] also showed similar results.

We particularly considered the situation that the filtering efficacy of masks being relatively low, and asked whether masks with intermediate efficacy in personal protection (for example, only filtering 50% of viruses) can have an aggregate effect when applied on a population-wide scale. In Fig S3, masks with different filtering efficacy are considered. The results show that even masks only trapping 20% of the virus can bring a significant impact when generally adopted by the population. Also, in the model shown in http://www.zhiyuanlab.xyz/MASK_0906.html, more details related to the effect of population-wide mask-wearing were considered, such as infections relying on non-respiratory routes, and the different filtering efficacies for masks as source control and personal protection.

### 4.5 Solution with imported cases

Border control measures can effectively stop imported cases of viral carriers. To examine the time needed for their effect to take place, let us first consider growth driven by imported cases when unchecked. Under a daily flux $J_{\mathrm{ext}}(t)$ of imported cases, Eq. (S4) is modified to,

$$\dot{A}_1 = -\alpha_{\mathrm{A}} A_1 + \int_{-\infty}^{t} K\left(t - t_1\right) A_1\left(t_1\right) dt_1 + J_{\mathrm{ext}}(t). \tag{S51}$$

Consider a simple situation where the imported cases grow exponentially with a rate $\lambda$, i.e. $J_{\mathrm{ext}}(t) = J_{\mathrm{I}} e^{\lambda_{\mathrm{I}} t}$. Let's seek an exponentially growing solution where the local population is driven by the imported cases,

$$\begin{aligned} A_1(t) &= A_{\mathrm{all}}\, e^{\lambda_{\mathrm{I}} t}, \\ \left(\lambda_{\mathrm{I}} + \alpha_{\mathrm{A}}\right) A_{\mathrm{all}} &= A_{\mathrm{all}}\, \tilde{K}\left(\lambda_{\mathrm{I}}\right) + J_{\mathrm{I}}, \\ A_{\mathrm{all}} &= \frac{J_{\mathrm{I}}}{\lambda_{\mathrm{I}} + \alpha_{\mathrm{A}} - \tilde{K}\left(\lambda_{\mathrm{I}}\right)}. \end{aligned}$$

The fraction of local infections after the initial transient is given by,

$$\frac{A_{\mathrm{local}}}{A_{\mathrm{all}}} = \frac{\tilde{K}\left(\lambda_{\mathrm{I}}\right)}{\lambda_{\mathrm{I}} + \alpha_{\mathrm{A}}}. \tag{S52}$$

In many cases, the ratio of imported and local infections is known. Equation (S52) can then be used to calibrate the level of local transmission when the imported cases grow exponentially at a rate greater than the one given in Table S2 for local outbreaks.

### 4.6 Crossover behaviour under linear transmission reduction

We consider here a situation where the disease transmission rate per infected individual in a given community gradually weakens according to the schedule,

$$\eta(t) = 1 - \left(1 - \eta_1\right) \frac{t - t_0}{T}, \quad t_0 < t < t_0 + T.$$

Here $t_0$ is the starting date and $T$ is the duration of the change. After this period, $\eta(t) = \eta_0$. Equation (S2) then takes the form,

$$\dot{A}_1 = -\alpha_\mathrm{A} A_1 + \int_{-\infty}^{t} K(t - t_1)\, \eta(t_1)\, A_1(t_1)\, dt_1. \tag{S53}$$

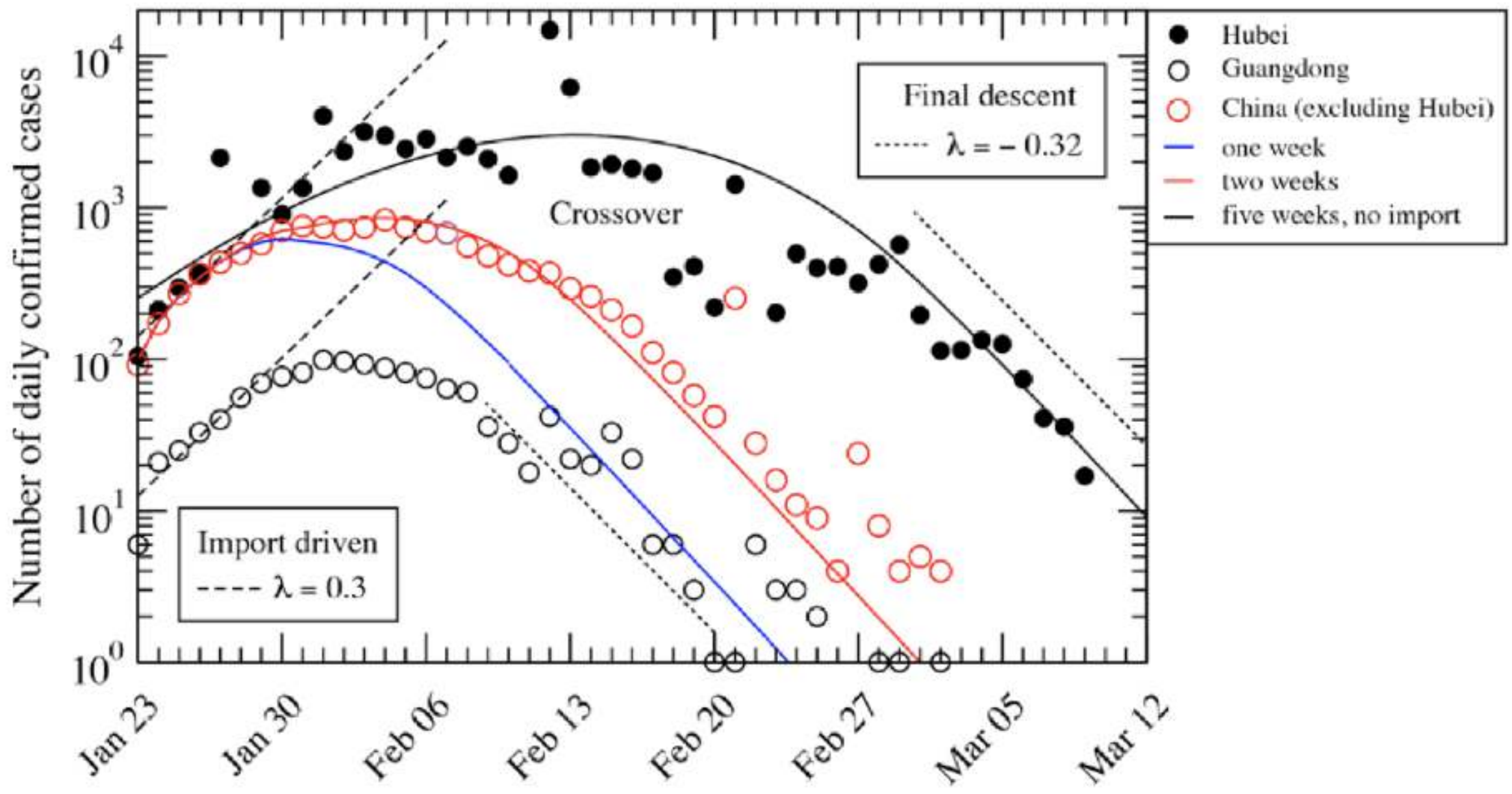

**Figure S4: Simulation of the number of daily confirmed cases with a linear decay of transmission per infected individual.** The number of daily confirmed cases in Hubei, Guangdong and China (excluding Hubei) are re-plotted for comparison. Note that the lockdown in Wuhan was started on the date Jan 23. Cluster infections result in occasional bursts in the time series which are not modeled in this work.

We used Eq. (S53) to simulate the epidemic development curves in China after the Wuhan lockdown (Figs. 4a and 4b, Main Text), taking $\eta_1 = 0$. The starting time of the social distancing policy $t_0$ is chosen to be one week after the lockdown. The blue, red and black curves in Fig. S4 correspond to three different values of $T$ given in the legend. The situation represented by blue and red curves is initially driven by imported cases whose number grows exponentially at a rate $\lambda_0 = 0.2$/day prior to the lockdown, with an amplitude that decreases linearly and vanishes on the 10th day after the lockdown. No imported cases were introduced to generate the black curve. Taken at face value, our model can reproduce fast or slow crossovers seen in the data from various provinces of China. On the scale of tens of millions of people, Eq. (S53) can only be considered as representing an aggregated trend helped by the linear nature of the model. As the epidemic spreads into disparate communities, there could be huge variations in the level of transmission from community to community, particularly as the policies tighten. Occasional outbreaks are evident from the scattered data in the declining phase of the pandemic.

### 4.7 Homestay

The situation on the cruise ship Diamond Princess is close to a sudden complete confinement of the passengers. Such a scenario is described by the schedule function

$$\eta(t) = \begin{cases} 1, & t < t_0 \\ \eta_1, & t > t_0 \end{cases}$$

at $\eta_1 = 0$. Under the Markovian assumption introduced above, the change over from exponential growth to exponential decay can be solved analytically.

Without loss of generality, we may define our time such that $t_0 = 0$. Take $A_1(t) = A_0 e^{\lambda_0 t}$ for $t < t_0 = 0$ , we may rewrite Eq. (S53) as,

$$\dot{A}_1 = -\alpha_A A_1 + \int_{-\infty}^{0} K(t-t_1) A_0 e^{\lambda_0 t_1} dt_1 + \eta_1 \int_0^t K(t-t_1) A_1(t_1) dt_1. \tag{S54}$$

Let

$$S(t) = A_0 \int_{-\infty}^{0} K(t-t_1) e^{\lambda_0 t_1} dt_1,$$

Eq. (S54) can be rewritten as,

$$\dot{A}_1 = -\alpha_A A_1 + S(t) + \eta_1 \int_0^t K(t-t_1) A_1(t_1) dt_1. \tag{S55}$$

Performing the Laplace transform of Eq. (S55), we obtain,

$$\tilde{A}_1(\lambda) = \frac{\tilde{S}(\lambda) + A_0}{\alpha_A + \lambda - \eta_1 \tilde{K}(\lambda)}, \tag{S56}$$

where

$$\tilde{S}(\lambda) = A_0 \frac{\tilde{K}(\lambda) - \tilde{K}(\lambda_0)}{\lambda - \lambda_0}.$$

Under Eq. (S45), we have

$$\tilde{K}(\lambda) = \tilde{K}(\lambda_0) \frac{\alpha_{L,\text{eff}} + \lambda_0}{\alpha_{L,\text{eff}} + \lambda} = (\alpha_A + \lambda_0) \frac{\alpha_{L,\text{eff}} + \lambda_0}{\alpha_{L,\text{eff}} + \lambda}.$$

Consequently,

$$\tilde{A}_1(\lambda) = A_0 \frac{\alpha_{L,\text{eff}} - \alpha_A + \lambda - \lambda_0}{(\alpha_A + \lambda)(\alpha_{L,\text{eff}} + \lambda) - \eta_1 (\alpha_A + \lambda_0)(\alpha_{L,\text{eff}} + \lambda_0)}. \tag{S57}$$

Carrying out inverse transform of Eq. (S57), we obtain

$$A_1(t) = A_0 \left( B_+ e^{\lambda_+ t} + B_- e^{\lambda_- t} \right). \tag{S58}$$

Here,

$$\lambda_\pm = \frac{-(\alpha_{L,\text{eff}} + \alpha_A) \pm \sqrt{(\alpha_{L,\text{eff}} - \alpha_A)^2 + 4 R_1 \alpha_{L,\text{eff}} \alpha_A}}{2},$$

$$B_+ = \frac{\lambda_+ - \lambda_0 + \alpha_{L,\text{eff}} - \alpha_A}{\lambda_+ - \lambda_-},$$

$$B_- = 1 - B_+,$$

with

$$R_1 = \eta_1 R_E = \eta_1 \frac{(\alpha_A + \lambda_0)(\alpha_{L,\text{eff}} + \lambda_0)}{\alpha_{L,\text{eff}} \alpha_A}.$$

The solution is shown for several examples in Fig. S5, where we assumed the "daily confirmed" cases to be proportional to $A_1(t)$. Note that the duration of the crossover is set by $\alpha_{L,\text{eff}}^{-1} \simeq 3$ days for the parameters chosen.

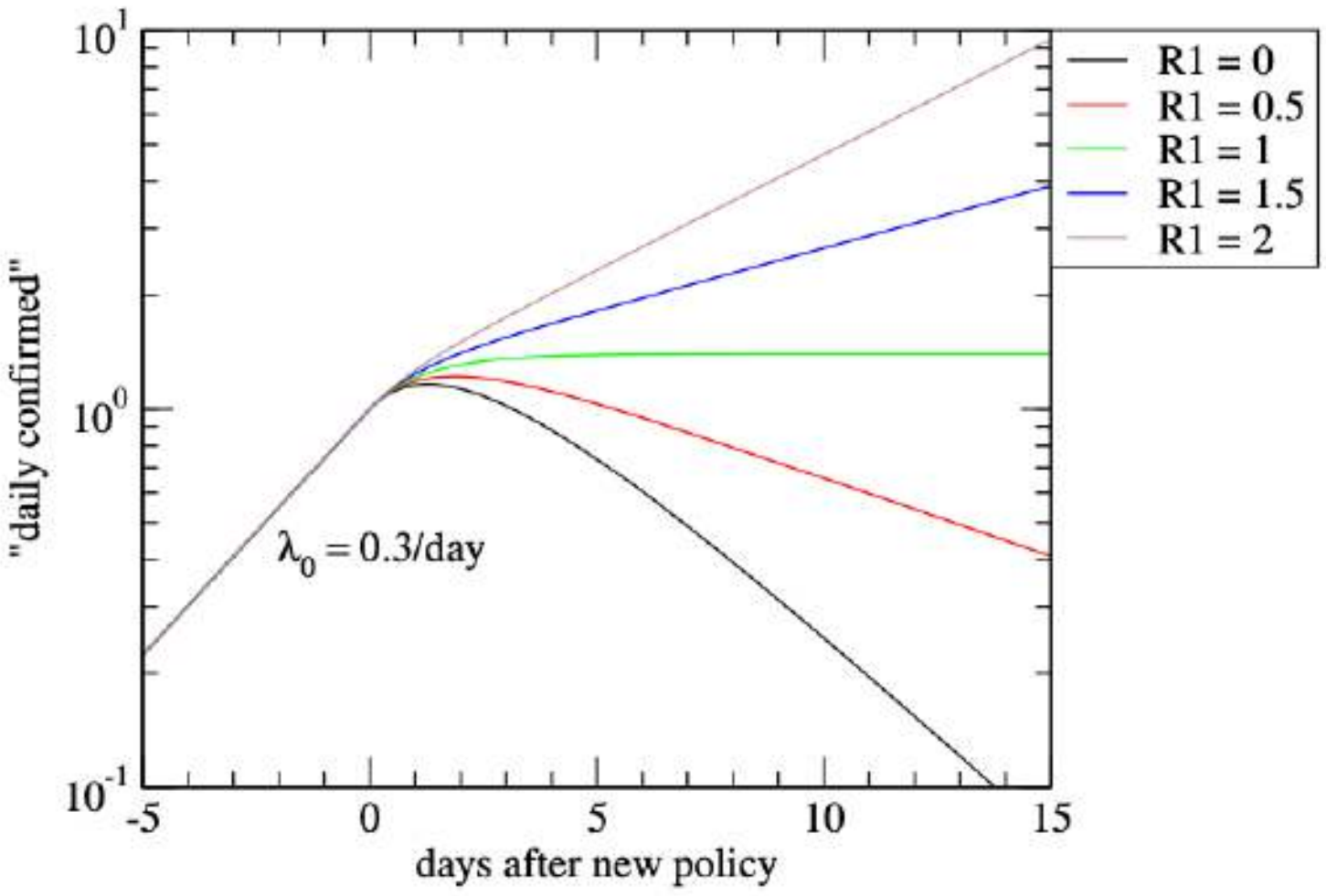

**Figure S5: Simulation of daily confirmed infections under a sudden reduction of** $R_{\rm E}$. The epidemic initially grows at a rate $\lambda_0 = 0.3/{\rm day}$ before the jump, corresponding to $R_{\rm E} = 3.8$. Different curves correspond to different values of $R_1$ under the new policy. Here $\alpha_{\rm L,eff} = 0.3/{\rm day}$, $\alpha_{\rm A} = 0.5/{\rm day}$.

## 5 Estimation of exponential growth rates during initial outbreaks

The reported COVID-19 infection cases in various countries and regions typically show exponential growth with time during the initial outbreak of the disease. The growth rate $\lambda$ is of great interest not only to epidemiologists, but also to the general public. In theory, $\lambda$ can be determined by fitting the cumulative number of confirmed cases to an exponential function of time (days). However, when the number of cases is small, growth tends to be affected by chance events (e.g., super-spreaders) and specific communities where the outbreak first took place. Furthermore, it may be dominated by imported cases which do not reflect much of the spreading characteristics in the local community. Therefore caution is required in using the data to extract the epidemic growth rate which itself may vary with time due to the mitigation and intervention measures introduced.

In their study of epidemic data for dengue fever in Brazil, Favier *et al.* [24] proposed to plot the daily new cases $\Delta N$ against the cumulative confirmed cases $N(t)$ and to fit the data with a linear function to determine $\lambda$. For perfect exponential growth, we have

$$\frac{\Delta N(t)}{N(t)} \equiv \frac{N(t) - N(t - \Delta t)}{N(t)} = 1 - \exp(-\lambda \Delta t), \tag{S59}$$

where $\Delta t = 1$ day. Therefore the slope of the scatter plot $\Delta N(t)$ against $N(t)$ can be used to estimate $\lambda$. The next step is to determine the interval over which the fitting is carried out.

Following the work by Favier *et al.*, we performed linear fits of $\Delta N(t)$ against $N(t)$, each time using one more data point, through which we obtain the slope and goodness-of-fit of these linear regressions as a function of cumulative number of confirmed cases. Then considering the 80th percentile of the slopes left to the minimum goodness-of-fit with goodness-of-fit smaller than 0.5 as the exponential phase, we obtain an estimate of the growth rate. The data and fitted linear functions are shown in Fig. S6.

In Table S10, we collect the estimated growth rates using the above scheme, along with the statistics of the linear regression (R-squared and p-value for the F-test) and also the total case number in the exponential growth phase for different countries/regions. Where applicable, the estimated values are in good agreement with the location of the plateaus in Fig. 6d. Given the many technical issues involved, we prefer not to include this particular analysis in the Main Text but instead have made revisions to the effect that the epidemic growth rate generally varied from place to place.

Solving for $\lambda$ in Eq. (S59), we obtain,

$$\lambda(t) = \frac{\ln N(t) - \ln N(t - \Delta t)}{\Delta t}. \tag{S60}$$

Therefore we may also think of the above procedure as computing the instantaneous growth rates from the $\ln N(t)$ against $t$ curve, with $\Delta t$ chosen suitably to minimize irregularities in testing and reporting. Given the mean incubation period of about 6 days for COVID-19, $\lambda(t)$ is expected to be a slow-varying quantity with a typical timescale of a week or longer. Five examples of $\lambda(t)$ versus $N(t)$ during the initial local outbreaks are given in Fig. 6d of the Main Text. Reasonably good agreement is seen between the plateau value of $\lambda(t)$ and the growth rates listed in Table S10 in four of the five countries, with the exception of Italy, where the substantially lower value of 0.26/day in the table corresponds to the second plateau shown in Fig. 6d.

Provincial epidemic growth in China after the Wuhan lockdown has a large component of imported cases. The epidemic development in the Hubei province, on the other hand, is subject to swift intervention and containment measures such as isolation of patients in Fangcang hospitals [25]. The situation in South Korea is also special due to a well-documented outbreak in Daegu in the second half of February [26]. Therefore the growth rates listed in Table S10 in these locations need to be interpreted differently.

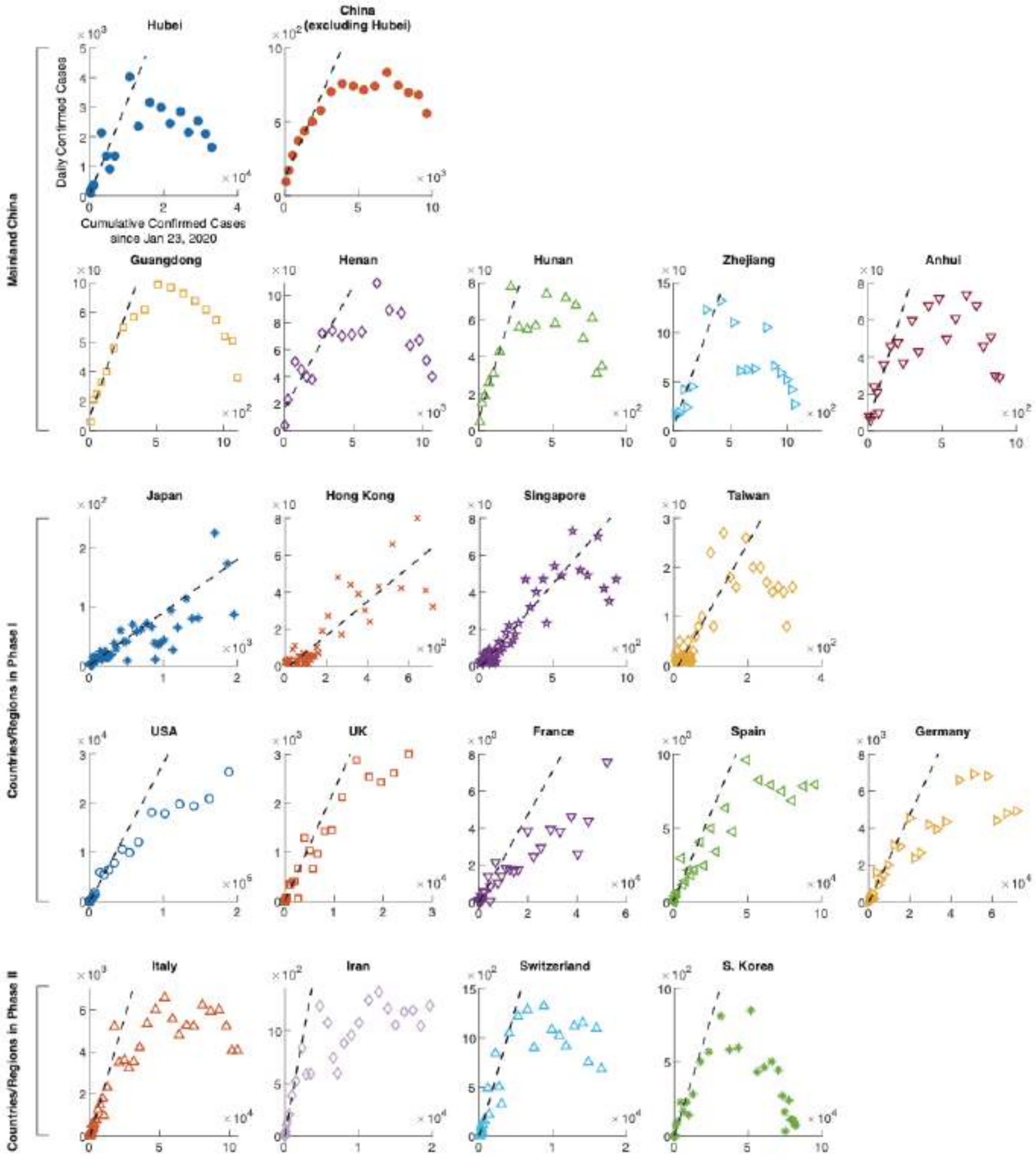

**Figure S6:** Linear regression of the number of daily confirmed cases as a function of the cumulative number of confirmed cases in the beginning of the COVID-19 pandemic in selected countries/regions.

**Table S10:** Estimated exponential growth rate $\lambda$, linear regression statistics (coefficient of determination $R^2$ and $p$-value for the $F$-test), and total case number in exponential growth phase since January 23, 2020 for different countries/regions, following the scheme of Favier *et al.* [24]. The calculated mean reproduction number $R_E$ is also shown (95% confidence interval shown in parentheses).

| Countries /Regions | $\lambda$ | $R^2$ | $p$-value ($F$-test) | Total case number in exponential phase | Mean reproduction number $R_E$ |
|---|---|---|---|---|---|
| Hubei | 0.3661 | 0.7678 | 0.0019 | 10,733 | 4.94 (4.21,5.89) |
| China (excluding Hubei) | 0.2571 | 0.9443 | 0.0012 | 1,851 | 3.28 (2.93,3.71) |
| Guangdong | 0.2975 | 0.9620 | 5.4775e-04 | 181 | 3.83 (3.37,4.42) |
| Henan | 0.1999 | 0.6634 | 0.0257 | 273 | 2.60 (2.38,2.87) |
| Hunan | 0.3004 | 0.9697 | 3.4891e-04 | 139 | 3.88 (3.40,4.48) |
| Zhejiang | 0.3985 | 0.9165 | 1.8789e-04 | 418 | 5.53 (4.64,6.70) |
| Anhui | 0.3079 | 0.7994 | 0.0066 | 151 | 3.99 (3.49,4.63) |
| Japan | 0.0934 | 0.7243 | 1.3134e-09 | 418 | 1.62 (1.55,1.70) |
| Hong Kong | 0.0975 | 0.6664 | 5.3861e-14 | 410 | 1.65 (1.58,1.73) |
| Singapore | 0.0960 | 0.7422 | 3.0786e-14 | 345 | 1.64 (1.57,1.72) |
| Taiwan | 0.1416 | 0.6959 | 4.2117e-11 | 168 | 2.02 (1.90,2.17) |
| USA | 0.3318 | 0.9398 | 4.2590e-21 | 25,488 | 4.36 (3.77,5.12) |
| UK | 0.2561 | 0.8222 | 7.2719e-11 | 5,018 | 3.27 (2.92,3.69) |
| France | 0.2701 | 0.7119 | 3.2394e-08 | 6,633 | 3.45 (3.07,3.93) |
| Spain | 0.2691 | 0.7469 | 2.1631e-07 | 9,942 | 3.44 (3.06,3.91) |
| Germany | 0.2715 | 0.8255 | 9.6580e-12 | 5,795 | 3.47 (3.08,3.95) |
| Italy | 0.2575 | 0.8755 | 5.7844e-11 | 17,660 | 3.29 (2.94,3.72) |
| Iran | 0.4821 | 0.9937 | 2.4192e-12 | 978 | 7.34 (5.91,9.21) |
| Switzerland | 0.3019 | 0.8068 | 4.1837e-07 | 2,700 | 3.90 (3.42,4.51) |
| S. Korea | 0.3858 | 0.8871 | 1.9155e-10 | 832 | 5.29 (4.47,6.37) |